\documentclass[journal]{IEEEtran}

\usepackage{amssymb}
\usepackage{amsmath}
\usepackage{cite}
\usepackage{url}
\usepackage{pifont}
\usepackage{empheq}
\usepackage{xcolor}
\usepackage{graphicx}
\usepackage{subfigure}
\usepackage{enumitem}
\usepackage{fancyhdr}
\usepackage{mdwmath}	
\usepackage{mdwtab}
\usepackage{caption}
\usepackage{amsthm}
\usepackage{gensymb}

\usepackage{algorithm}
\usepackage{algorithmic}

\newtheorem{lemma}{Lemma}
\newtheorem{remark}{Remark}
\newtheorem{theorem}{Theorem}
\newtheorem{corollary}{Corollary}
\newtheorem{assumption}{Assumption}
\newtheorem{definition}{Definition}
\newtheorem{proposition}{Proposition}

\newcommand{\eqr}[1]{(\ref{#1})}
\newcommand{\fref}[1]{Fig.~\ref{#1}}

\newcommand*{\QEDA}{\null\nobreak\hfill\ensuremath{\blacksquare}}

\begin{document}
\title{Generalized Pinching-Antenna Systems:\\ A Leaky-Coaxial-Cable Perspective}
\author{Kaidi~Wang,~\IEEEmembership{Member,~IEEE,}
Zhiguo~Ding,~\IEEEmembership{Fellow,~IEEE,}
and~Lajos~Hanzo,~\IEEEmembership{Life~Fellow,~IEEE}
\thanks{K. Wang is with the Department of Electrical and Electronic Engineering, the University of Manchester, Manchester, M1 9BB, UK (email: kaidi.wang@ieee.org).}
\thanks{Z. Ding is with the School of Electrical and Electronic Engineering (EEE), Nanyang Technological University, Singapore 639798 (e-mail: zhiguo.ding@ntu.edu.sg).}
\thanks{L. Hanzo is with the School of Electronics and Computer Science, University of Southampton, Southampton, SO17 1BJ, UK (e-mail: lh@ecs.soton.ac.uk).}}
\maketitle
\begin{abstract}
The evolution toward the sixth-generation (6G) wireless networks has flexible reconfigurable antenna architectures capable of adapting their radiation characteristics to the surrounding environment. At the center stage, while waveguide based pinching antennas have been shown to beneficially ameliorate wireless propagation environments, their applications have remained confined to high-frequency scenarios. As a remedy, we propose a downlink generalized pinching-antenna system that adapts this compelling concept to low-frequency operation through a leaky-coaxial-cable (LCX) implementation. By endowing LCX structures with controllable radiation slots, the system inherits the key capabilities of waveguide based pinching antennas. Explicitly, these include reconfigurable line-of-sight (LoS) links, reduced path loss, and flexible deployment, while supporting a practical implementation of the pinching-antenna concept at low frequencies. A twin-stage propagation model is developed for characterizing both the guided transmission and wireless radiation encountered over LoS and non-line-of-sight (NLoS) paths. Analytical results reveal strong local gain, complemented by rapid distance-dependent decay. Hence, we conceive a matching joint optimization framework, which maximizes throughput by harnessing game theoretic association and convex power allocation. Simulation results demonstrate substantial performance gains over conventional fixed-antenna benchmarks.
\end{abstract}
\section{Introduction}
The evolution toward the next-generation (NG) wireless networks has witnessed a paradigm shift in antenna design, emphasizing the need for flexible reconfigurable antenna architectures capable of dynamically adapting their radiation characteristics to the surrounding environment. Over the past few years, several emerging technologies have followed this trend, including reconfigurable intelligent surfaces (RISs), fluid antennas, and movable antennas \cite{wu2019irs, wong2020fluid, zhu2023modeling}. RISs are capable of wavefront manipulation by adjusting the phase responses of nearly passive reflecting elements, thereby enhancing signal propagation. However, their two-hop cascaded link structure inevitably introduces additional path loss and erodes efficiency \cite{wu2021intelligent, yang2026movable}. By contrast, fluid and movable antennas exploit spatial diversity by altering antenna positions within a constrained region \cite{wu2024fluid, ma2024movable}. Although these technologies improve propagation conditions through position reconfigurability, their reconfiguration region is typically confined to a wavelength-scale local space around the transceiver \cite{wu2026fluid, lai2026fluid, wu2026fluid2, liu2026movable}.

To overcome these limitations, the pinching-antenna concept has recently been proposed \cite{suzuki2022pinching, ding2024pin}. Unlike conventional antennas that radiate from fixed ports, a pinching antenna relies on a dielectric waveguide, where electromagnetic leakage can be generated at arbitrary positions along the waveguide by locally ``pinching'' the guiding medium \cite{liu2025pinching}. This unique mechanism facilitates large scale adjustment of the effective antenna position and convenient reconfiguration of the radiation pattern without requiring additional active radio-frequency chains \cite{kaidi2025pin, yang2025pinching}. The resultant flexibility allows dynamic reshaping of radiation points to circumvent physical blockages and arrange for line-of-sight (LoS) links in previously obstructed environments, thereby significantly improving propagation reliability \cite{ding2025blockage, kaidi2025pin4}. Furthermore, the  pinching-antenna structure can be realized using low-cost dielectric materials and simple mechanical control, which makes it an attractive candidate for scalable and energy-efficient reconfigurable antenna deployments \cite{xu2025generalized}. As a result, pinching antennas provide precise spatial adaptability, enhanced link robustness, and low implementation cost, circumventing the fundamental limitations of RISs, fluid antennas, and movable antennas.
\subsection{State-of-the-Art}
Extensive research efforts have been focused on investigating the theory, optimization, and applications of pinching-antenna systems \cite{wang2025pa, tyrovolas2025pin, ouyang2025uplink, xie2025pin, kaidi2025pin2, xu2025pin2, mao2025pin, kaidi2025pin3}. The electromagnetic modeling and beamforming characteristics of pinching-antenna systems were first studied in \cite{wang2025pa}, where a coupled mode theoretical framework was established to characterize the radiation mechanism and guide joint transmit as well as pinching beamforming design. Building on this foundation, \cite{tyrovolas2025pin} developed a comprehensive analytical framework that incorporated both free-space attenuation and waveguide loss, deriving closed-form expressions for outage probability and average rate, quantifying the influence of waveguide attenuation on system performance. To further address scalability and practical deployment, \cite{ouyang2025uplink} proposed a segmented pinching-antenna architecture that divides long dielectric waveguides into shorter sections, mitigating inter-antenna coupling and propagation loss while providing an efficient solution for uplink transmission modeling.

In parallel, optimization oriented research has advanced the system design of pinching-antenna aided wireless communications. The authors of \cite{xie2025pin} proposed a low-complexity algorithm for optimizing the specific positions and user association, providing closed-form designs for both orthogonal and non-orthogonal multiple access (OMA/NOMA). A practical implementation was investigated in \cite{kaidi2025pin2}, where discrete pre-installed pinching antennas were considered as a realistic hardware architecture. By activating or deactivating a subset of pre-installed antennas, this approach substantially simplifies the antenna placement optimization, while maintaining flexible beamforming capability. Similarly, \cite{xu2025pin2} formulated a quality-of-service (QoS)-aware NOMA framework and developed iterative optimization algorithms for jointly adjusting the antenna positions and power allocation, achieving an improved balance between throughput and user fairness. 

Beyond conventional communication scenarios, recent studies have extended the concept of pinching antennas to emerging NG paradigms. The authors of \cite{mao2025pin} introduced a multi-waveguide pinching-antenna based integrated sensing and communication architecture, where joint beamforming and antenna deployment were optimized for striking a trade-off between radar sensing accuracy and communication throughput. Furthermore, \cite{kaidi2025pin3} explored the use of phase adjustment in pinching antennas to enhance physical layer security. By exploiting the controllable phase shifts among multiple pinching antennas, a coalitional game based activation algorithm was proposed, demonstrating that cooperative phase alignment significantly improves the secrecy performance compared to conventional fixed-antenna based systems.
\subsection{Motivation and Contributions}
Although the above studies have established a solid foundation for pinching-antenna assisted wireless systems, they have primarily focused on high-frequency scenarios, where dielectric waveguides exhibit low propagation loss and compact form factors. By contrast, the application of pinching antennas in low-frequency environments\footnote{In this paper, the term low-frequency primarily refers to sub-6 GHz bands, in contrast to the high-frequency regimes commonly considered in dielectric waveguide based pinching-antenna studies.} remains largely unexplored, mainly due to the excessive physical dimensions and material constraints of dielectric waveguides at longer wavelengths. Nevertheless, low-frequency wireless systems are essential for NG applications such as industrial Internet-of-Things, underground communication, and high-reliability indoor connectivity, where robust propagation, spatial flexibility, and cost efficiency are required.

Recent studies on generalized pinching-antenna systems have suggested that the pinching-antenna principle can be extended to lower-frequency transmission media \cite{xu2025generalized}. In this context, leaky coaxial cables (LCXs) constitute a suitable candidate for lower-frequency implementations, as they provide a well-established guided transmission platform with practical deployment advantages. Motivated by this insight, we investigate the feasibility of realizing the generalized pinching-antenna principle through an LCX based implementation. Inspired by the principles of pre-deployed pinching antennas \cite{kaidi2025pin, kaidi2025pin2}, the proposed system selectively activates or deactivates radiation slots to enable on-demand control of leakage positions. This mechanism transforms the conventional LCX philosophy of fixed radiation points into a reconfigurable LCX based pinching-antenna concept that preserves the spatial flexibility of waveguide based pinching antennas, while retaining the robustness and practicality of coaxial-cable structures. Consequently, this architecture supports adaptive field distribution along the cable and enables spatially flexible and energy-efficient communication in indoor or tunnel-like environments.

The main contributions are summarized as follows:
\begin{itemize}
\item A novel LCX based generalized pinching-antenna system is proposed, enabling spatially flexible electromagnetic radiation control through slot activation and deactivation. A comprehensive twin-stage propagation model is established for jointly characterizing guided transmission within the cable and wireless radiation toward users over both LoS and non-line-of-sight (NLoS) paths. The model incorporates an angle-dependent radiation term that captures the directional characteristics of LCX radiation, demonstrating its inherent interference suppression capability and enhanced spatial selectivity.
\item A detailed analytical study is conducted to characterize the achievable data rate of the LCX based generalized pinching-antenna system. Our closed-form expressions derived for the single-user single-slot case provide insights into its spatial power distribution and propagation behavior. The analysis reveals that the LCX based channel exhibits strong local gain complemented by rapid distance-dependent decay, leading to effective interference suppression in multi-cable scenarios. These observations confirm that LCX structures can be utilized to realize generalized pinching-antenna functionality.
\item In the LCX based generalized pinching-antenna system, the elevation angle and the associated user-slot distance play a dominant role in determining the channel strength, leading to fundamentally different association behavior. Because the geometric advantage of reduced distance can outweigh intra-cable interference, multiple users may be assigned to the same cable. This geometry-dependent effect introduces new challenges and motivates a sum rate maximization problem that jointly considers user assignment, slot activation, and power allocation.
\item The mixed-integer problem formulated is decoupled into a pair of tractable subproblems. The first subproblem, involving user assignment and slot activation, is modeled as a coalitional game in which each cable acts as a game player. A coalition formation algorithm is developed to jointly determine the user-to-cable associations and slot activation states. The second subproblem, corresponding to power allocation, is reformulated as a convex optimization framework, which is efficiently solved using the successive convex approximation method.
\end{itemize}
Simulation results validate the effectiveness of the proposed LCX based generalized pinching-antenna system. The results demonstrate that the proposed architecture achieves significant improvements both in sum rate and outage probability compared to conventional fixed-antenna systems. Moreover, the proposed joint optimization framework effectively enhances the overall system throughput while guaranteeing the individual rate requirements of all users. These results confirm the advantages of employing LCX based pinching antennas for low-frequency reconfigurable wireless communication systems.
\section{System Model}
\begin{figure}[!t]
\centering{
\subfigure[]{\centering{\includegraphics[width=90mm]{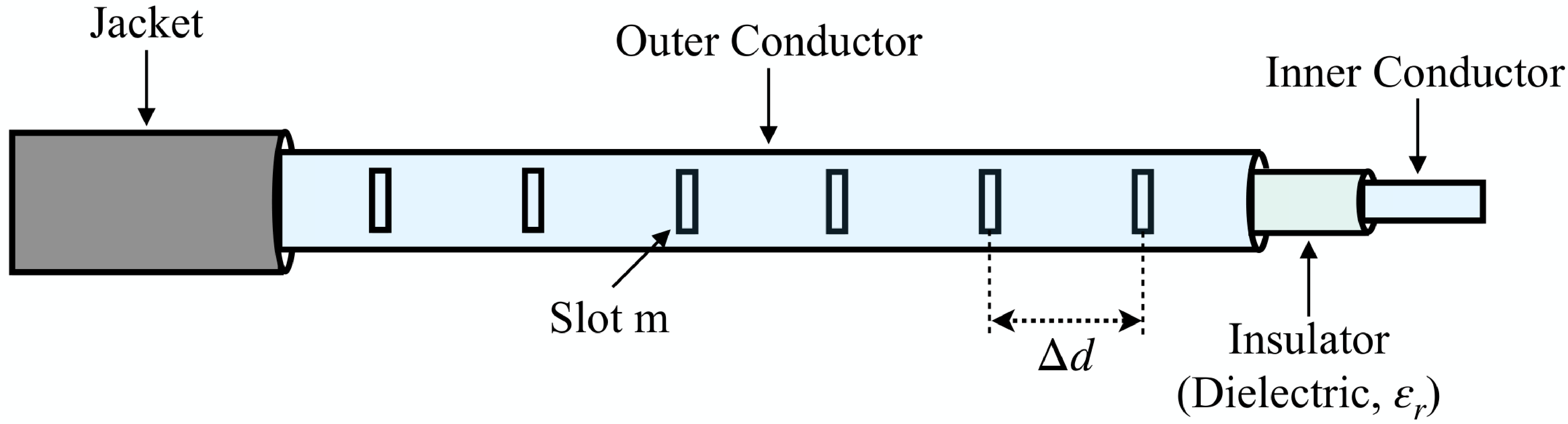}}}
\subfigure[]{\centering{\includegraphics[width=90mm]{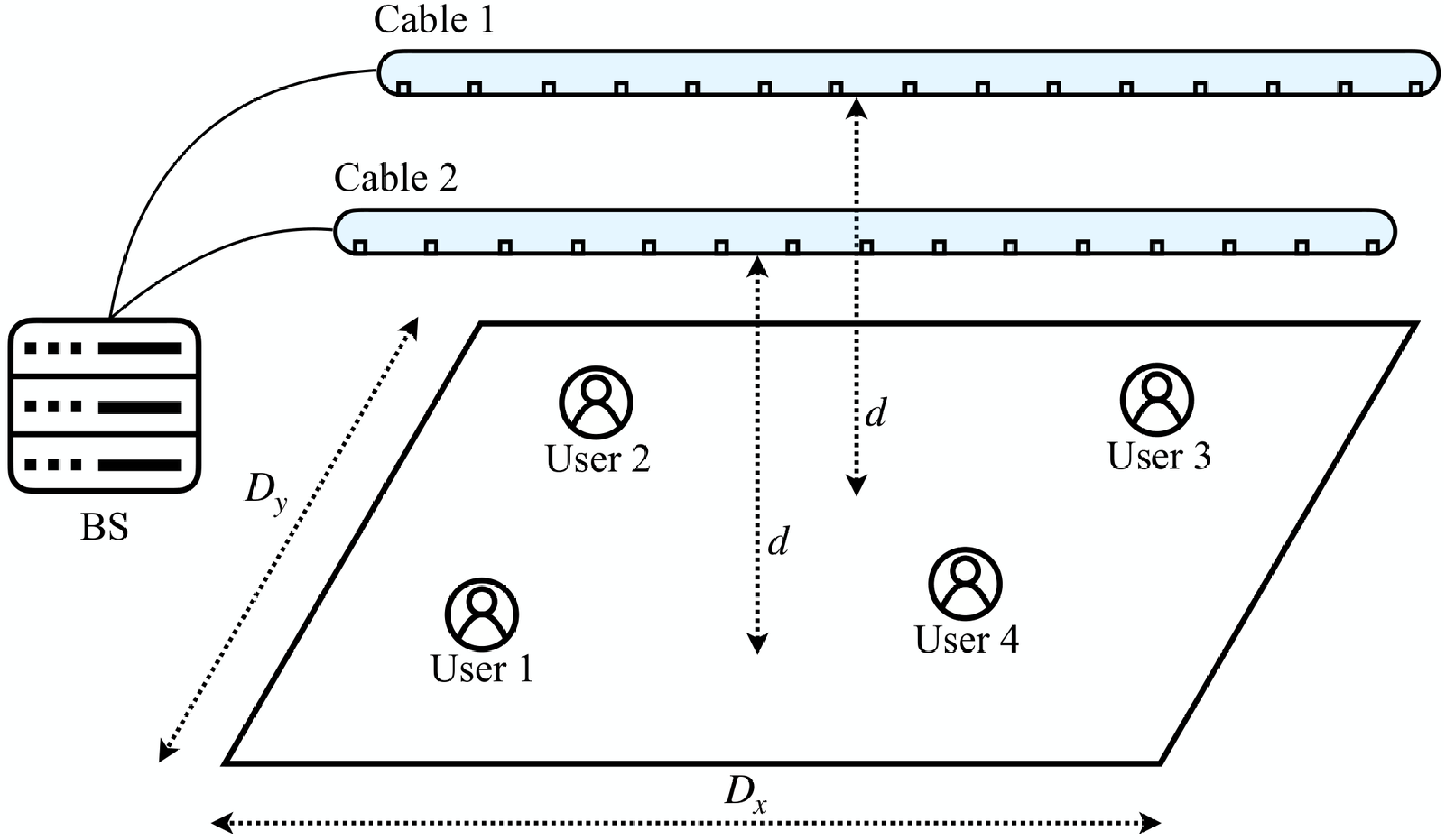}}}}
\caption{An illustration of the proposed LCX based generalized pinching-antenna system.}
\label{system}
\end{figure}

Consider a downlink LCX based generalized pinching-antenna system employed in a rectangular region of dimensions $D_x \times D_y$, as illustrated in \fref{system}. Within this region, $K$ coaxial cables are installed in parallel at height $d$, each aligned with the $x$-axis and spanning the entire length of $D_x$\footnote{In practical deployment, the $K$ LCXs are fixed to the ceiling using standard brackets or clamps placed at intervals of approximately $0.5$-$1$~m, ensuring parallel alignment and mechanical stability along $D_x$ \cite{myllymaki2019leaky}. Their lateral separation is determined by the region width $D_y$ and the number of cables, such that the LCXs are uniformly spaced and centered within the coverage area. Moreover, consistent with common practice, the radiating slots are oriented downward toward the user region, as illustrated in \fref{system}(b).}. Each cable contains $M$ periodically spaced vertical radiating slots, which act as pre-installed pinching antennas and simultaneously serve $N$ users. The index sets of cables, users, and slots on cable $k$ are denoted by $\mathcal{K}=\{1,2,\cdots, K\}$, $\mathcal{N}=\{1,2,\cdots, N\}$, and $\mathcal{M}_k=\{1,2,\cdots,M\}$, respectively. For ease of reference, the main symbols used throughout this paper are summarized in Table~\ref{notation}.

\begin{table}[t]
\centering
\caption{Summary of Main Symbols}\vspace{-1mm}
\label{notation}
\begin{tabular}{cl}
\hline
Symbol & Definition \\ \hline
$D_x$, $D_y$ & Length and width of the service region \\
$d$ & Deployment height of the cables \\
$\Delta d$ & Spacing between adjacent slots on each cable \\
$K$, $N$, $M$ & Numbers of cables, users, and slots per cable \\
$\mathcal{K}$, $\mathcal{N}$, $\mathcal{M}_k$ & Sets of cables, users, and slots on cable $k$ \\
$\boldsymbol{\psi}_n$ & Position of user $n$ \\
$\boldsymbol{\psi}_{k,0}$ & Feed point position of cable $k$ \\
$\boldsymbol{\psi}_{k,m}^\mathrm{slot}$ & Position of slot $m$ on cable $k$ \\
$\boldsymbol{\psi}_{\ell}^\mathrm{scat}$ & Position of scatterer $\ell$\\
$\kappa$, $\varepsilon_r$ & Cable attenuation constant and relative permittivity\\
$f_c$, $\lambda$ & Carrier frequency and free-space wavelength\\
$h_{k,m,n}$ & Channel to user $n$ via slot $m$ on cable $k$ \\
$h_{k,n}$ & Channel from cable $k$ to user $n$ \\
$\phi_{k,m,n}$ & Elevation angle between slot $m$ on cable $k$ and user $n$ \\
$\alpha_{k,n}$ & User assignment indicator between cable $k$ and user $n$ \\
$\beta_{k,m}$ & Activation indicator of slot $m$ on cable $k$ \\
$p_{k,n}$ & Power allocation coefficient for user $n$ on cable $k$ \\
$P_t$, $\sigma^2$ & Available transmit power and noise power\\
$N_c$ & Number of cables serving users \\
$N_k$ & Number of activated slots on cable $k$ \\
$R_n$ & Achievable data rate of user $n$ \\
$R_{\min}$ & Minimum target data rate of each user \\
$\mathcal{A}$, $\mathcal{B}$ & User assignment and slot activation coalitions \\
\hline
\end{tabular}
\vspace{-3mm}
\end{table}

In the system considered, users are randomly distributed over the ground plane, and the position of user $n$ is denoted by $\boldsymbol{\psi}_n=(x_n,y_n,0)$, where $x_n\in[-\tfrac{D_x}{2},\tfrac{D_x}{2}]$ and $y_n\in[-\tfrac{D_y}{2},\tfrac{D_y}{2}]$. All coaxial cables are connected to a base station (BS), and the feed point of cable $k$ is given by $\boldsymbol{\psi}_{k,0}=(x_k,y_k,d)$, where $x_k=-\tfrac{D_x}{2}$ and $y_k=-\tfrac{D_y}{2}+(k-\tfrac{1}{2})\tfrac{D_y}{K}$. For each cable, the distance between two adjacent slots is denoted by $\Delta d$, with $\Delta d=\tfrac{D_x}{M-1}$. Accordingly, the center position of the $m$-th slot on cable $k$ is expressed as $\boldsymbol{\psi}_{k,m}^\mathrm{slot}=(x_m,y_k,d)$, where $x_m=-\tfrac{D_x}{2}+(m-1)\Delta d$.
\subsection{Channel Model}
In the LCX based generalized pinching-antenna system proposed, the downlink transmission consists of two consecutive phases, including the guided propagation within the coaxial cables and the wireless propagation between the radiating slots and users. During guided propagation, the signal experiences both attenuation and phase variation along the transmission path \cite{torrance1996lcx, hou2023lcx}. Accordingly, the channel spanning from the feed point of cable $k$ to its $m$-th slot can be expressed as follows:
\begin{equation}\label{cableh}
h_{k,m}^\mathrm{cable}=10^{-\frac{\kappa}{20}\|\boldsymbol{\psi}_{k,0}-\boldsymbol{\psi}_{k,m}^\mathrm{slot}\|}e^{-j\frac{2\pi}{\lambda}\sqrt{\varepsilon_r}\|\boldsymbol{\psi}_{k,0}-\boldsymbol{\psi}_{k,m}^\mathrm{slot}\|},
\end{equation}
where $\kappa$ (in dB/m) is the longitudinal attenuation constant of the coaxial cable, $\lambda$ is the free-space wavelength, $\varepsilon_r$ is the relative permittivity of LCX, and $\|\boldsymbol{\psi}_{k,0}-\boldsymbol{\psi}_{k,m}^\mathrm{slot}\|$ is the distance between the feed point and the $m$-th slot of cable $k$.

\begin{figure}[!t]
\centering{\includegraphics[width=85mm]{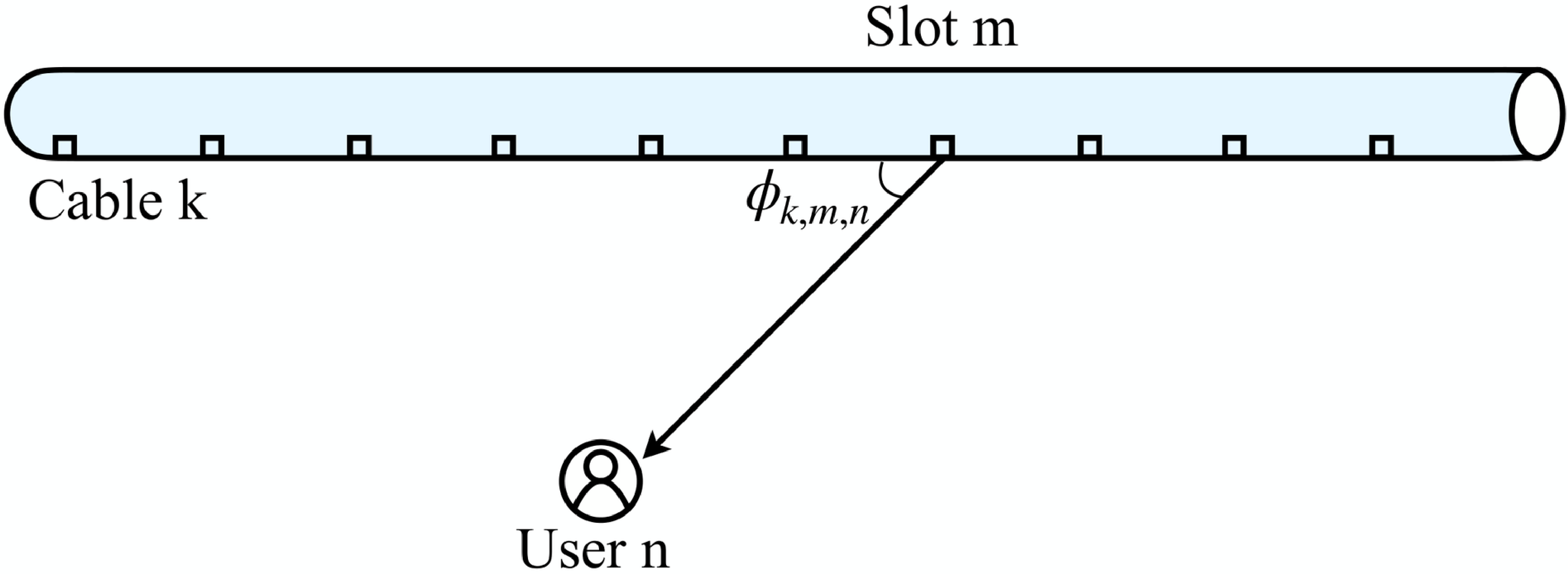}}
\caption{An illustration of the elevation angle $\phi_{k,m,n}$ between the $m$-th slot on cable $k$ and user $n$.}
\label{angle}
\end{figure}

In the wireless transmission phase, each slot functions as a magnetic dipole that radiates electromagnetic energy toward the users. The radiated field amplitude decays with distance and depends on the radiation direction of the slot. Each engineered slot can therefore be regarded as an equivalent radiator whose main lobe is directed downward, i.e., toward the user plane~\cite{morgan1999lcx, wang2001lcx}. According to electromagnetic theory, the field amplitude radiated by a small-loop slot toward the elevation angle $\phi$ is proportional to $\sin(\phi)$~\cite{yin2024lcx}. As a result, the direct channel between user $n$ and the $m$-th slot on cable $k$ can be formulated as follows:
\begin{equation}\label{radh}
h_{k,m,n}^\mathrm{rad}=\eta\frac{e^{-j\frac{2\pi}{\lambda}\|\boldsymbol{\psi}_n-\boldsymbol{\psi}_{k,m}^\mathrm{slot}\|}}{\| \boldsymbol{\psi}_n-\boldsymbol{\psi}_{k,m}^\mathrm{slot}\|}\sin(\phi_{k,m,n}),
\end{equation}
where $\eta=\tfrac{c}{4\pi f_c}$, $c$ is the speed of light, $f_c$ is the carrier frequency, $\|\boldsymbol{\psi}_n-\boldsymbol{\psi}_{k,m}^\mathrm{slot}\|$ is the distance between user $n$ and the $m$-th slot on cable $k$, and $\phi_{k,m,n}$ is the corresponding elevation angle. As shown in \fref{angle}, $\phi_{k,m,n}$ is defined as the angle between the line connecting the $m$-th slot on cable $k$ and user $n$, and the horizontal plane, i.e., $\sin(\phi_{k,m,n})=\tfrac{d}{\|\boldsymbol{\psi}_n-\boldsymbol{\psi}_{k,m}^\mathrm{slot}\|}$.

For the NLoS link, multiple scatterers are considered to construct the reflected propagation paths, where the signal radiated from each slot impinges on the scatterers and is subsequently reflected toward the users \cite{zhang2022channel}. The NLoS link between the $m$-th slot on cable $k$ and user $n$ is modeled as
\begin{equation}\label{scath}
h_{k,m,n}^\mathrm{scat}\!\!=\!\eta\!\sum_{\ell=1}^L\!\frac{\delta_\ell e^{-j\frac{2\pi}{\lambda}\!\left(\|\boldsymbol{\psi}_\ell^\mathrm{scat}\!-\boldsymbol{\psi}_{k,m}^\mathrm{slot}\|+\|\boldsymbol{\psi}_n\!-\boldsymbol{\psi}_\ell^\mathrm{scat}\|\right)}}{\|\boldsymbol{\psi}_\ell^\mathrm{scat}\!-\boldsymbol{\psi}_{k,m}^\mathrm{slot}\|\|\boldsymbol{\psi}_n\!-\boldsymbol{\psi}_\ell^\mathrm{scat}\|}\!\sin(\phi_{k,m,\ell}),
\end{equation}
where $L$ is the number of scatterers, $\delta_\ell$ is the complex gain of the $\ell$-th scattering path, $\boldsymbol{\psi}_\ell^\mathrm{scat}\!=\!(x_\ell,y_\ell,z_\ell)$ is the position of scatterer $\ell$, and $\phi_{k,m,\ell}$ is the angle between the line connecting the $m$-th slot on cable $k$ and scatterer $\ell$.

Based on~\eqr{cableh} and~\eqr{radh}, the LoS channel from the feed point of cable $k$ to user $n$ via the $m$-th slot is given by
\begin{equation}\label{los}
h_{k,m,n}^\mathrm{LoS}=h_{k,m}^\mathrm{cable}h_{k,m,n}^\mathrm{rad}=\frac{\eta A_{k,m}e^{-jB_{k,m,n}}}{\|\boldsymbol{\psi}_n-\boldsymbol{\psi}_{k,m}^\mathrm{slot}\|}\sin(\phi_{k,m,n}),
\end{equation}
where
\begin{equation}
A_{k,m}=10^{-\frac{\kappa}{20}\|\boldsymbol{\psi}_{k,0}-\boldsymbol{\psi}_{k,m}^\mathrm{slot}\|},
\end{equation}
and
\begin{equation}
B_{k,m,n}=\frac{2\pi}{\lambda}\sqrt{\varepsilon_r}\|\boldsymbol{\psi}_{k,0}-\boldsymbol{\psi}_{k,m}^\mathrm{slot}\|+\frac{2\pi}{\lambda}\|\boldsymbol{\psi}_n-\boldsymbol{\psi}_{k,m}^\mathrm{slot}\|.
\end{equation}

Similarly, the NLoS channel can be obtained from~\eqr{cableh} and~\eqr{scath}, which is given by
\begin{align}
h_{k,m,n}^\mathrm{NLoS}&=h_{k,m}^\mathrm{cable}h_{k,m,n}^\mathrm{scat}\\\nonumber
&=\eta A_{k,m}\!\!\sum_{\ell=1}^L\!\frac{\delta_\ell e^{-jC_{k,m,\ell,n}}}{\|\boldsymbol{\psi}_\ell^\mathrm{scat}\!-\!\boldsymbol{\psi}_{k,m}^\mathrm{slot}\|\|\boldsymbol{\psi}_n\!-\!\boldsymbol{\psi}_\ell^\mathrm{scat}\|}\!\sin(\phi_{k,m,\ell}),
\end{align}
where we have:
\begin{align}
C_{k,m,\ell,n}&=\frac{2\pi}{\lambda}\sqrt{\varepsilon_r}\|\boldsymbol{\psi}_{k,0}-\boldsymbol{\psi}_{k,m}^\mathrm{slot}\|\\\nonumber
&\quad+\frac{2\pi}{\lambda}\!\left(\|\boldsymbol{\psi}_\ell^\mathrm{scat}-\boldsymbol{\psi}_{k,m}^\mathrm{slot}\|+\|\boldsymbol{\psi}_n\!-\boldsymbol{\psi}_\ell^\mathrm{scat}\|\right).
\end{align}

By integrating these two propagation phases, the overall channel between the feed point of cable $k$ and user $n$ through the $m$-th slot can be presented as follows:
\begin{equation}\label{channel}
h_{k,m,n}=h_{k,m,n}^\mathrm{LoS}+h_{k,m,n}^\mathrm{NLoS}.
\end{equation}
In this work, perfect geometry based channel state information (CSI) is assumed to be available at the BS for the proposed optimization. Specifically, the channel coefficients are determined by the positions of users, slots, cables, and scatterers, together with the cable and propagation parameters. Accordingly, the proposed framework is mainly applicable to static or slowly varying scenarios, where the geometry based CSI and the corresponding user association and slot activation decisions remain valid over a sufficiently long time interval.

According to the channel model seen in~\eqr{channel}, several observations can be obtained.
\begin{remark}\label{re1}
In the LCX based generalized pinching-antenna system proposed, the elevation angle $\phi$ between any user and slot satisfies $\phi \in (0,90^\circ]$, leading to $\sin(\phi) \in (0,1]$. This angle-dependent radiation behavior is an inherent property of LCX slots, whereas conventional waveguide based pinching-antenna structures do not exhibit this characteristic. The resultant directionality enhances spatial focusing and interference management, thereby making the LCX based design particularly suitable for low-frequency implementations.
\end{remark}

\begin{remark}
Due to the presence of the $\sin(\phi)$ term in \eqr{channel}, the received power is proportional to $d^2/(\rho^2+d^2)^2$, where $\rho=\sqrt{(x_n-x_m)^2+(y_n-y_k)^2}$ is the horizontal offset. Hence, the height $d$ serves as a deployment parameter that balances angular gain against path-length loss. For a fixed $d$, this term is strictly decreasing in $\rho$ and reaches its maximum at $\rho=0$, corresponding to the user being directly beneath the slot. For a given $\rho>0$, it is unimodal in $d$ and attains its unique maximum at $d=\rho$. In practical deployments, adjusting $d$ to the environmental geometry can further enhance performance.
\end{remark}

\begin{remark}
The proposed framework can be readily extended to the uplink. In the uplink transmission, each LCX slot acts as a receiving aperture whose reception pattern is reciprocal to its downlink radiation pattern. Accordingly, similar to the downward conical radiation region in the downlink, each slot primarily captures signals arriving within a conical region directed toward the cable. Therefore, the same angular factor of $\sin(\phi_{k,n})$ appears in the uplink channel gain.
\end{remark}
\subsection{Signal Model}
In the LCX based generalized pinching-antenna system considered, each user is assumed to be served by a single cable. Specifically, the signals of all users are available at the BS, and each cable only transmits the desired signals of its assigned users. In this context, user assignment is introduced, where a binary indicator $\alpha_{k,n}\in\{0,1\}$ is employed to denote the association between user $n$ and cable $k$. As a result, the superposed signal transmitted by cable $k$ can be expressed as follows:
\begin{equation}
x_k=\sum_{n=1}^N\sqrt{p_{k,n}}\alpha_{k,n}s_n,
\end{equation}
where $p_{k,n}$ is the power allocation coefficient for the signal of user $n$ transmitted through cable $k$, and $s_n$ is the desired signal of user $n$.

To enable adaptive control of the radiated field distribution along the LCX, a slot activation model is adopted, which can be physically interpreted using reconfigurable slot blocking structures \footnote{Prior LCX related studies have conceptually considered switching individual slots between radiating and non-radiating states to generate distinct channel patterns \cite{nagayama2022lcx1, nagayama2022lcx2}, while related patent designs have explored practical slot control mechanisms \cite{asplund2016leaky}. Motivated by these concepts, a feasible realization is to integrate a small controllable shutter or an electronic actuating element at each slot, allowing the aperture to be covered or exposed and thereby deactivating or activating the corresponding slot.}. Consequently, the channel between the feed point of cable $k$ and user $n$ can be modeled as the superposition of magnetic dipoles representing the activated slots, as follows:
\begin{equation}\label{channeln}
h_{k,n}=\sum_{m=1}^M\beta_{k,m}h_{k,m,n},
\end{equation}
where $\beta_{k,m}\in\{0,1\}$ is the slot activation indicator. Specifically, $\beta_{k,m}=0$ indicates that the $m$-th slot on cable $k$ is deactivated, whereas $\beta_{k,m}=1$ denotes activation. It is worth noting that no independent per slot phase precoding is assumed in \eqr{channeln}. The phase shifts induced by guided and wireless propagation are incorporated into $h_{k,m,n}$, and the summation in \eqr{channeln} represents complex valued channel superposition. Under the narrowband equivalent channel model adopted, path length differences are captured as phase shifts, while inter-symbol interference is assumed to be negligible. Based on this slot activation mechanism, the following remark provides insights into its effect on system performance. 

\begin{remark}
Through slot activation, the radiated power in downlink transmission can be concentrated toward regions having higher communication demand, thereby enhancing the local signal-to-noise ratio (SNR) and achievable data rate. In uplink transmission, slot activation can be employed to suppress signals from unassigned users, thus reducing interference and improving individual data rate.
\end{remark}

It is assumed that the total transmit power at the BS is evenly distributed among all cables serving at least one user, while any cable having no assigned user is ignored. Accordingly, a coefficient $\tfrac{1}{N_\mathrm{c}}$ is introduced, where $N_\mathrm{c}=\sum_{k=1}^K\mathbf{1}_{\{\sum_{n=1}^N\alpha_{k,n}\neq 0\}}$ is the number of cables serving users, and $\mathbf{1}_{\{\cdot\}}$ is the indicator function. Moreover, since all slots exhibit identical electromagnetic characteristics, the available transmit power of each cable is equally shared among its activated antennas, leading to another coefficient $\tfrac{1}{N_k}$, where $N_k=\max\{1, \sum_{m=1}^M\beta_{k,m}\}$ is the number of activated slots on cable $k$. As a result, the signal received at user $n$ can be expressed as follows:
\begin{align}
y_n&=\sqrt{\frac{P_t}{N_\mathrm{c}}}\sum_{k=1}^K \sqrt{\frac{1}{N_k}}h_{k,n}x_k+w_n\\\nonumber
&=\sqrt{\frac{P_t}{N_\mathrm{c}}}\sum_{k=1}^K\sqrt{\frac{1}{N_k}}\sum_{m=1}^M\beta_{k,m}h_{k,m,n}\sum_{i=1}^N\sqrt{p_{k,i}}\alpha_{k,i}s_i+w_n,
\end{align}
where $P_t$ is the total transmit power, and $w_n$ is the additive noise at user $n$. At the receiver, each user detects its desired signal by treating all other signals as interference, and the corresponding achievable data rate is given by
\begin{equation}\label{rate}
R_n=\log_2\!\!\left(\!1\!+\!\frac{\frac{P_t}{N_\mathrm{c}}\sum_{k=1}^K\frac{1}{N_k}p_{k,n}\alpha_{k,n}|h_{k,n}|^2}{\frac{P_t}{N_\mathrm{c}}\!\sum_{k=1}^K\!\frac{1}{N_k}\!\sum_{i=1,i\neq n}^Np_{k,i}\alpha_{k,i}|h_{k,n}|^2\!+\!\sigma^2}\!\right),
\end{equation}
where $\sigma^2$ is the noise power. When multiple users share the same cable, successive interference cancellation (SIC) can be incorporated as an optional receiver enhancement to reduce intra-cable interference. However, SIC is not employed in this work, and each user treats unintended signals as interference, as shown in \eqr{rate}. Alternatively, the proposed framework is applicable to OMA by assigning orthogonal resources to users, thereby eliminating inter-user interference at the cost of reduced resource utilization.

Based on the above power allocation and slot activation model, the following remark provides additional insights.
\begin{remark}
For any user $n$ assigned to cable $k$, the received contribution is given by $\tfrac{1}{N_k}\big|\sum_{m=1}^{M}\beta_{k,m}h_{k,m,n}\big|^2$, where $N_k$ is the number of activated slots. Activating more slots enables constructive multi-slot combining, but simultaneously reduces the power allocated to each slot. Hence, there exists a fundamental trade-off between the spatial combining gain and power dilution, which determines the overall received signal strength.
\end{remark}
\subsection{Data Rate Analysis}
\begin{figure}[!t]
\centering{\includegraphics[width=90mm]{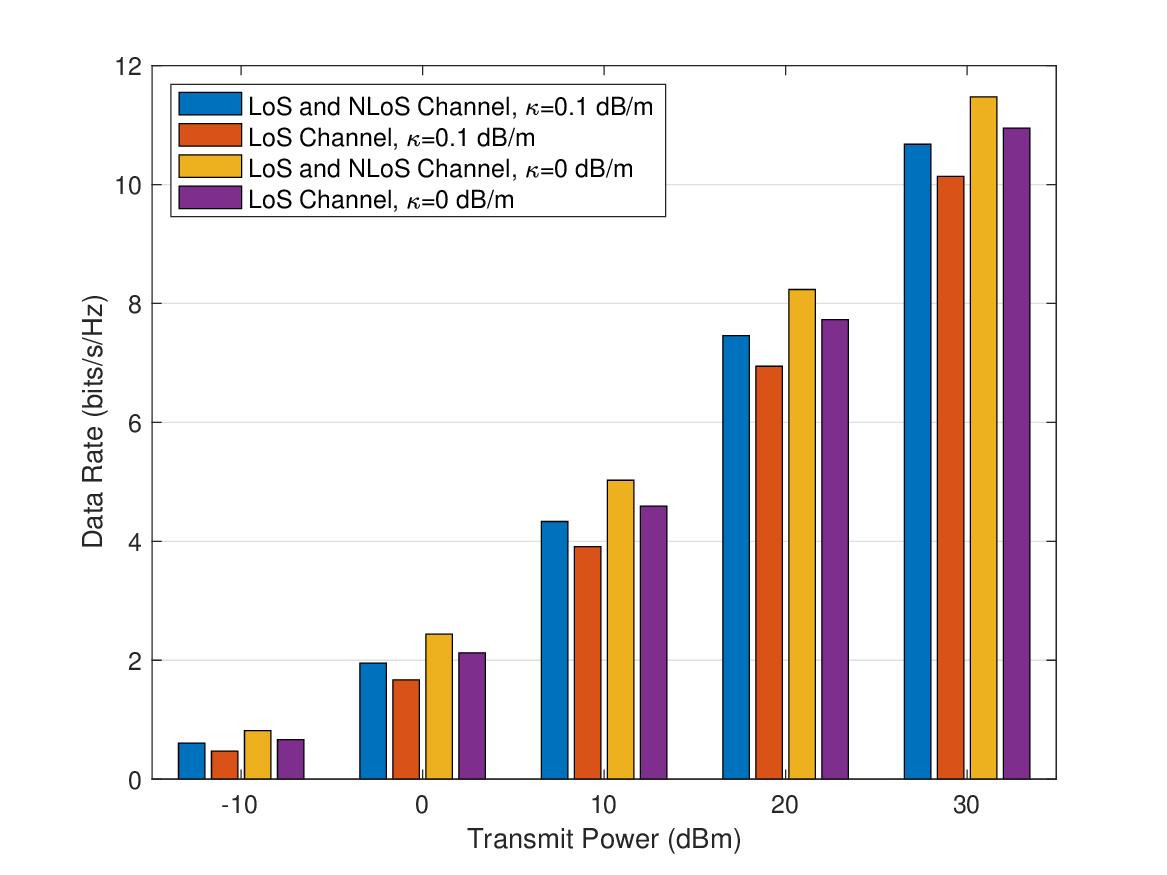}}
\caption{Influence of NLoS propagation and cable attenuation, where $D_x=50$,$D_y=30$~m, $K=1$, $N=1$, $M=50$, and $L=10$. The closest slot is activated.}
\label{impact}
\end{figure}

This subsection focuses on the analysis of the achievable data rate in the LCX based generalized pinching-antenna system proposed. To this end, a simple case with a single user served by a single activated slot is considered. As shown in \fref{impact}, since the impact of the NLoS path component and the attenuation in cables is negligible, these effects are ignored in the subsequent analysis. Moreover, because only one slot is activated, the phase term has no effect, hence it is omitted. Under these conditions, the data rate of user $n$ is expressed as follows:
\begin{equation}
R_n^\mathrm{LCX}=\log_2\left(1+\frac{P_n\eta^2d^2}{\sigma^2\|\boldsymbol{\psi}_n-\boldsymbol{\psi}_{k,m}^\mathrm{slot}\|^4}\right),
\end{equation}
where $P_n$ is the transmit power of the desired signal for user $n$. To enhance the data rate, an intuitive strategy is to assign each user to the closest cable and activate the nearest slot. For comparison, a conventional fixed-antenna system is also considered, in which a single antenna is installed at $\boldsymbol{\psi}_0=(0,0,d)$. The corresponding data rate of user $n$ is given by
\begin{equation}
R_n^\mathrm{fix}=\log_2\left(1+\frac{P_n\eta^2}{\sigma^2\|\boldsymbol{\psi}_n-\boldsymbol{\psi}_0\|^2}\right).
\end{equation}

The following remarks characterize the spatial distribution properties and power scaling behavior of the proposed LCX based architecture, as compared to its conventional fixed-antenna counterpart.
\begin{remark}
For randomly distributed users, the proposed LCX based generalized pinching-antenna system can be regarded as a network associated with user positions relative to the nearest slot/cable distributed within $x \in [-\tfrac{\Delta_x}{2}, \tfrac{\Delta_x}{2}]$ and $y \in [-\tfrac{\Delta_y}{2}, \tfrac{\Delta_y}{2}]$, where $\Delta_x=\Delta d$ and $\Delta_y=D_y/K$. By contrast, the conventional system corresponds to a distribution over $x \in [-\tfrac{D_x}{2}, \tfrac{D_x}{2}]$ and $y \in [-\tfrac{D_y}{2}, \tfrac{D_y}{2}]$.
\end{remark}

\begin{remark}
In the single-slot case, the received power scales as $|h|^2\propto\eta^2 d^2/r^4$ with $r=\|\boldsymbol{\psi}_n-\boldsymbol{\psi}_{k,m}^\mathrm{slot}\|$, whereas for a conventional fixed antenna it scales as $|h|^2\propto\eta^2/r^2$. Hence, the LCX based link provides strong local gain near the active slot, but decays more rapidly with distance. This property can be exploited to effectively mitigate interference and enhance physical layer security.
\end{remark}

Building upon the above observations, the following proposition compares the achievable data rates of the proposed LCX based and conventional fixed-antenna systems.
\begin{proposition}\label{compare1}
At high SNR, the proposed LCX based generalized pinching-antenna system outperforms the conventional fixed-antenna system in terms of its achievable data rate, if the following condition holds:
\begin{equation}
\left(1\!+\!\frac{D^2}{4d^2}\right)^{\left(1\!+\!\frac{4d^2}{D^2}\right)}\ge e\!\left(1\!+\!\frac{\Delta_x^2\!+\!\Delta_y^2}{4d^2}\right)^2.
\end{equation}
Furthermore, the data rate difference between these two systems increases monotonically with $\tfrac{D}{d}$, where $D=\min\{D_x, D_y\}$.
\end{proposition}
\begin{IEEEproof}
Refer to Appendix~A.
\end{IEEEproof}

From Proposition~\ref{compare1}, the following remark can be obtained:
\begin{remark}
In the LCX based generalized pinching-antenna system proposed, since $\Delta_x=\Delta d$ and $\Delta_y=D_y/K$, reducing the inter-slot spacing $\Delta d$ and/or increasing the number of cables $K$ can further improve the overall system performance.
\end{remark}

As discussed in Remark~\ref{re1}, the presence of the $\sin(\phi)$ term in \eqr{channel} implies that, in the LCX based generalized pinching-antenna channel model having a single activated slot, a user can only achieve the same data rate as that in the conventional pinching-antenna channel model, when it is located directly beneath the active slot. In the multi-cable scenario, however, the directional radiation pattern of the LCX structure effectively mitigates inter-cable interference, thereby enhancing the achievable data rate. This observation leads to the following proposition.
\begin{proposition}\label{compare2}
Consider the single-slot activation case in which user $n$ is served by slot $m$ on LCX $k$ and experiences interference from slot $m'$ on LCX $k'$. Assuming that all users transmit with equal power, the LCX based pinching-antenna channel model achieves an individual data rate no lower than that of the conventional pinching-antenna channel model if and only if
\begin{equation}
\sin^2(\phi_{k,m,n}) \ge \frac{\gamma_{k',m',n}+\sin^2(\phi_{k',m',n})}{1+\gamma_{k',m',n}},
\end{equation}
where
\begin{equation}
\gamma_{k',m',n}\triangleq\frac{\sigma^2}{\eta^2}\|\boldsymbol{\psi}_n-\boldsymbol{\psi}_{k',m'}^\mathrm{slot}\|^2 10^{\frac{\kappa}{10}\|\boldsymbol{\psi}_{k',0}^\mathrm{slot}-\boldsymbol{\psi}_{k',m'}^\mathrm{slot}\|}.
\end{equation}
In the high-SNR regime $(\sigma^2 \to 0)$, this condition simplifies to
\begin{equation}
\|\boldsymbol{\psi}_n-\boldsymbol{\psi}_{k,m}^\mathrm{slot}\|\le \|\boldsymbol{\psi}_n-\boldsymbol{\psi}_{k',m'}^\mathrm{slot}\|,
\end{equation}
or equivalently,
\begin{equation}
\phi_{k,m,n}\ge \phi_{k',m',n}.
\end{equation}
\end{proposition}
\begin{IEEEproof}
Refer to Appendix~B.
\end{IEEEproof}

This observation can be intuitively interpreted as follows:
\begin{remark}
Proposition~\ref{compare2} reveals that the LCX based channel model achieves a higher instantaneous rate than the conventional waveguide based pinching-antenna model, whenever the serving slot provides a larger elevation angle or, equivalently, it is located closer to the user than the interfering slot. This condition reflects the inherent geometric advantage introduced by LCX slot radiation.
\end{remark}
\section{Problem Formulation}
Based on the LCX based generalized pinching-antenna channel model developed, an optimization problem is formulated in this section to characterize the achievable performance limits of the proposed architecture. Again, our design objective is to maximize the overall sum rate by optimally configuring the slot activation pattern and transmission power, while satisfying the individual QoS constraints. The problem is expressed as follows:
\begin{subequations}
\begin{empheq}{align}
\max_{\boldsymbol{\alpha},\boldsymbol{\beta},\mathbf{p}}\quad & \sum_{n=1}^NR_n\\
\textrm{s.t.} \quad & \alpha_{k,n}\in\{0,1\},\forall n\in\mathcal{N},\forall k\in\mathcal{K},\\
& \beta_{k,m}\in\{0,1\},\forall m\in\mathcal{M}_k,\forall k\in\mathcal{K},\\
& \sum\nolimits_{k=1}^K\alpha_{k,n}=1, \forall n\in\mathcal{N},\\
& \sum\nolimits_{m=1}^M\beta_{k,m}\ge\alpha_{k,n}, \forall n\in\mathcal{N}, \forall k\in\mathcal{K},\\
& 0\le p_{k,n}\le \alpha_{k,n}, \forall n\in\mathcal{N}, \forall k\in\mathcal{K},\\
& \sum\nolimits_{n=1}^Np_{k,n}\le 1, \forall k\in\mathcal{K},\\
& R_n \ge R_\mathrm{min}, \forall n\in\mathcal{N},
\end{empheq}
\label{problem}
\end{subequations}\vspace{-2mm}\\
where $\boldsymbol{\alpha}$, $\boldsymbol{\beta}$, and $\mathbf{p}$ are the sets of user assignment indicators, slot activation indicators, and power allocation coefficients, respectively. Constraints~(\ref{problem}d) and~(\ref{problem}e) ensure that each user is assigned to a single cable, and that each cable with assigned users activates at least one slot. Constraint~(\ref{problem}f) specifies the feasible range of power allocation coefficients, where the coefficient for any cable without assigned users is zero. Moreover, constraint~(\ref{problem}h) enforces a minimum data rate no lower than $R_\mathrm{min}$.

Problem~\eqr{problem} is a mixed-integer programming problem, which generally cannot be solved directly. To facilitate efficient optimization, it is decoupled into two subproblems. By grouping all integer variables into one subproblem, the user assignment and slot activation problem can be presented as follows:
\begin{subequations}
\begin{empheq}{align}
\max_{\boldsymbol{\alpha},\boldsymbol{\beta}}\quad & \sum_{n=1}^NR_n\\\nonumber
\textrm{s.t.} \quad & \text{(\ref{problem}b)}, \text{(\ref{problem}c)}, \text{(\ref{problem}d)}, \text{and}\!\!\quad\!\!\! \text{(\ref{problem}e)}.
\end{empheq}
\label{uaaaproblem}
\end{subequations}\vspace{-2mm}\\
The power allocation problem is then given by
\begin{subequations}
\begin{empheq}{align}
\max_{\mathbf{p}}\quad & \sum_{n=1}^NR_n\\\nonumber
\textrm{s.t.} \quad & \text{(\ref{problem}f)}, \text{(\ref{problem}g)}, \text{and}\!\!\quad\!\!\!\text{(\ref{problem}h)}.
\end{empheq}
\label{paproblem}
\end{subequations}\vspace{-2mm}\\
The QoS constraint is incorporated into the power allocation subproblem, since the achievable data rate of each user directly depends on the allocated transmit power. Hence, satisfying the minimum rate requirement $R_\mathrm{min}$ is primarily governed by the power distribution across users.
\section{Two-Stage Optimization Algorithm Design}
The sum rate maximization problem formulated is of mixed-integer and non-convex nature, making direct optimization intractable. To address this, a two-stage framework is developed. In the first stage, a game theoretic method jointly optimizes user assignment and slot activation. In the second stage, a successive convex approximation (SCA) based algorithm refines the power allocation to further enhance the system throughput while satisfying the individual target rate requirements.
\subsection{Game Based User Assignment and Slot Activation}
This subsection focuses on the user assignment and slot activation problem formulated in \eqr{uaaaproblem}. To address the inherent integer programming challenge, a game theoretic approach is adopted, in which user assignment and slot activation are modeled as a pair of coalitional games. Specifically, each cable forms two coalitions that represent the assigned users and activated slots, respectively. By constructing the optimal coalition partitions for all cables, the user assignment and slot activation problem can be effectively solved.

For user assignment, a collection of coalitions $\mathcal{A}=\{A_1, A_2, \cdots, A_K\}$ is defined, where coalition $A_k$ represents the set of users assigned to cable $k$. According to constraints~(\ref{problem}b) and~(\ref{problem}d), the conditions $A_k\cap A_{k'}=\emptyset,\forall k\neq k'$ and $\cup_{k\in\mathcal{K}}A_k=\mathcal{N}$ must hold. In the user assignment game, each user can switch from one coalition to another based on his/her individual preference, expressed as follows:
\begin{equation}\label{userpref}
(A_k,\mathcal{A})\prec_n (A_{k'},\mathcal{A}') \Leftrightarrow \sum_{n=1}^N v_n(\mathcal{A}) <\sum_{n=1}^N v_n(\mathcal{A}'),
\end{equation}
where the utility of user $n$ in coalition $A_k$ is defined as
\begin{equation}
v_n(\mathcal{A})=\log_2\!\!\left(\!1\!+\!\frac{\frac{P_t}{N_\mathrm{c}}\frac{1}{N_k}p_{k,n}|h_{k,n}|^2}{\frac{P_t}{N_\mathrm{c}}\!\sum_{j=1}^K\!\frac{1}{N_j}\!\sum_{i\in A_j,i\neq n}p_{j,i}|h_{j,n}|^2\!+\!\sigma^2}\!\right),
\end{equation}
and $N_\mathrm{c}$ is rewritten as $N_\mathrm{c}=\sum_{k=1}^K \mathbf{1}_{\{A_k \neq \emptyset\}}$. This preference relationship implies that user $n$ currently in coalition $A_k$ prefers to join another coalition $A_{k'}$ (with $k \neq k'$) if such a move increases the overall system utility. Consequently, the coalition structure of the user assignment game is updated as follows:
\begin{equation}
\mathcal{A'}=\mathcal{A}\backslash\{A_k,A_{k'}\}\cup\{A_k\backslash n,A_{k'}\cup n\}.
\end{equation}
That is, the coalition structure is transformed from $\mathcal{A}$ to $\mathcal{A}'$ as a result of user $n$ switching from cable $k$ to cable $k'$.

With a given coalition structure of user assignment, the slot activation game is performed for all cables with assigned users. To this end, a coalition structure $\mathcal{B}=\{B_1,B_2,\cdots,B_k\}$ is defined, where $B_k$ denotes the set of activated slots on cable $k$. According to constraint~(\ref{problem}e), any cable with assigned users must activate at least one slot, i.e., $|B_k|\ge 1$ for all $|A_k|\ge 1$. In the slot activation game, a slot can be activated or deactivated, if such an action leads to a strict increase in the overall sum rate, which can be expressed as follows:
\begin{equation}\label{antennaact}
(B_k,\mathcal{B}) \prec_m (B_k\cup m,\mathcal{B}')\Leftrightarrow \sum_{n=1}^N v_n(\mathcal{B})<\sum_{n=1}^N v_n(\mathcal{B}'),
\end{equation}
and
\begin{equation}\label{antennadeact}
(B_k,\mathcal{B}) \prec_m (B_k\backslash m,\mathcal{B}')\Leftrightarrow \sum_{n=1}^N v_n(\mathcal{B})< \sum_{n=1}^N v_n(\mathcal{B}').
\end{equation}
These preference relationships indicate that the $m$-th slot on cable $k$ can be activated or deactivated by joining or leaving the corresponding coalition $B_k$, respectively. As a result, the coalition structure is updated from $\mathcal{B}$ to $\mathcal{B}'$, where $\mathcal{B}'=\mathcal{B}\backslash B_k\cup\{B_k\cup m\}$ for activation or $\mathcal{B}'=\mathcal{B}\backslash B_k\cup\{B_k\backslash m\}$ for deactivation.

\begin{algorithm}[t]
\caption{Coalitional Game based User Assignment and Slot Activation Algorithm}
\label{cgalg}
\begin{algorithmic}[1]
\STATE \textbf{Initialization}
\STATE Assign each user to the nearest cable to obtain $\mathcal{A}$
\STATE Activate the nearest slot for each user to obtain $\mathcal{B}$
\STATE \textbf{Main Loop}
\FOR{each user $n\in\mathcal{N}$ with current cable $k$}
\FOR{each $k'\in\mathcal{K}$ with $k'\neq k$}
\FOR{each cable $i\in\mathcal{K}$ with $|A_i|\neq 0$}
\FOR{each slot $m\in\mathcal{M}_i$}
\IF{$m\notin B_i$ \AND $(B_i,\mathcal{B})\prec_m (B_i\cup m,\mathcal{B}')$}
\STATE Activate slot $m$ and update $\mathcal{B}$ as $\mathcal{B}'$
\ELSE
\IF{$m\in B_i$ \AND $|B_i|>1$ \AND $(B_i,\mathcal{B})\prec_m (B_i\setminus m,\mathcal{B}')$}
\STATE Deactivate slot $m$ and update $\mathcal{B}$ as $\mathcal{B}'$
\ENDIF 
\ENDIF
\ENDFOR
\ENDFOR
\IF{$(A_k,\mathcal{A})\prec_n (A_{k'},\mathcal{A}')$}
\STATE Assign user $n$ to cable $k$ and update $\mathcal{A}$ as $\mathcal{A}'$
\ENDIF
\ENDFOR
\ENDFOR
\end{algorithmic}
\end{algorithm}

According to the defined preferences of users and slots, a coalitional game based user assignment and slot activation algorithm is proposed in Algorithm~\ref{cgalg}. In the algorithm conceived, users select cables based on their preferences, and slot activation is performed whenever the user assignment structure changes. Activating multiple slots on the same cable may introduce a phase alignment problem, as the radiated signals from different slots can combine either constructively or destructively \cite{kaidi2025pin3}. This effect can be exploited to suppress interference or to enhance the desired signal strength. However, identifying the optimal activation pattern for effective phase alignment is highly challenging. To obtain a near-optimal solution, slot activation is re-executed for all cables whenever any user changes his/her cable preference, as shown in lines~7-17. The main loop of the algorithm terminates when no user or slot changes occur within a complete iteration, after which the variables $\boldsymbol{\alpha}$ and $\boldsymbol{\beta}$ are obtained from $\mathcal{A}$ and $\mathcal{B}$, respectively. It is worth noting that Algorithm~\ref{cgalg} is executed centrally at the BS. Thus, user assignment and slot activation/deactivation are evaluated based on the geometry based CSI available at the BS, rather than through distributed information exchange among users. The main dynamic overhead is therefore associated with updating user locations and broadcasting the resultant assignment and slot activation decisions.

\subsubsection{Convergence}
In Algorithm~\ref{cgalg}, the user assignment and slot activation games are updated sequentially, and the proposed algorithm converges for arbitrary initializations. Specifically, each accepted user assignment or slot activation/deactivation update is performed only if it strictly increases the sum rate. Since the number of feasible user assignment and slot activation states is finite, the iterative procedure must terminate after a finite number of updates. To accelerate convergence, the nearest cable and nearest slot initialization policy is adopted.

\subsubsection{Stability}
The final coalition structures obtained from Algorithm~\ref{cgalg} satisfy Nash stability \cite{han2012game}, defined as follows:
\begin{definition}
In the proposed coalitional game, the coalition structures $\mathcal{A}$ and $\mathcal{B}$ are Nash stable if and only if
\begin{enumerate}
\item $(A_k,\mathcal{A})\succeq_n(A_{k'},\mathcal{A}'), \forall n\in A_k, \forall k'\neq k$, and 
\item $(B_k,\mathcal{B})\succeq_m(B_{k'},\mathcal{B}'), \forall B_k\cap B_{k'}\neq\emptyset, B_k \triangle B_{k'}=\{m\}$.
\end{enumerate}
\end{definition}
According to the above definition, in the final coalition structures, no user or slot has an incentive to change its current state, which is consistent with the termination condition of the proposed algorithm. Therefore, the solution obtained by Algorithm~\ref{cgalg} is guaranteed to be Nash stable. However, Nash stability does not necessarily imply global optimality. Hence, the proposed coalitional game based algorithm should be regarded as a suboptimal approach.

\subsubsection{Complexity}
The computational complexity of Algorithm~\ref{cgalg} is evaluated under the worst-case scenario, where all users can switch among all cables and every slot may be activated or deactivated. In this case, a total of $N K^2 M$ calculations are executed within a single iteration. Considering $C$ iterations in the main loop, the overall complexity can be expressed as $\mathcal{O}(C N K^2 M)$. 
\subsection{SCA based Power Allocation}
In this section, the results of user assignment and slot activation are assumed to be obtained, and the power allocation is optimized based on the given structure. From \eqr{rate}, the data rate of user $n$ can be expressed as the difference of two concave functions, as follows:
\begin{align}\nonumber
R_n&=\log_2\left(\sum_{k=1}^Kg_{k,n}\!\sum_{i=1}^N\alpha_{k,i}p_{k,i}\!+\!\sigma^2\right)\\
&\quad-\log_2\left(\sum_{k=1}^Kg_{k,n}\!\!\!\sum_{i=1,i\neq n}^N\!\!\!\!\alpha_{k,i}p_{k,i}\!+\!\sigma^2\right),
\end{align}
where we have:
\begin{equation}
g_{k,n}\triangleq\frac{P_t}{N_\mathrm{c}N_k}|h_{k,n}|^2.
\end{equation}

By introducing the slack variables $\mu_n$ and $\nu_n$, the original power allocation problem can be reformulated as follows:
\begin{subequations}
\begin{empheq}{align}
\max_{\mathbf{p},\boldsymbol{\mu},\boldsymbol{\nu}}\quad & \sum_{n=1}^N\left[\log_2(\mu_n)-\log_2(\nu_n)\right]\\
\textrm{s.t.} \quad &\sum_{k=1}^Kg_{k,n}\!\sum_{i=1}^N\alpha_{k,i}p_{k,i}\!+\!\sigma^2\!\ge\! \mu_n,\forall n\!\in\!\mathcal{N},\\
&\sum_{k=1}^Kg_{k,n}\!\!\sum_{i=1,i\neq n}^N\!\!\!\!\alpha_{k,i}p_{k,i}\!+\!\sigma^2\!\le\! \nu_n,\forall n\!\in\!\mathcal{N},\\
& \mu_n \ge \nu_n 2^{R_\mathrm{min}}, \forall n\in\mathcal{N}\\\nonumber
& \text{(\ref{problem}f)}, \text{and}\!\!\quad\!\!\! \text{(\ref{problem}g)},
\end{empheq}
\label{paproblem1}
\end{subequations}\vspace{-2mm}\\
where $\boldsymbol{\mu}$ and $\boldsymbol{\nu}$ are the sets of slack variables. Since the objective function is a sum of difference of convex functions, problem~\eqr{paproblem1} is non-convex. To address this challenge, the concave term $-\log_2(\nu_n)$ is approximated using a first-order Taylor expansion of $\log_2(\nu_n)$ at the point $\nu_n^{(t)}$, given by
\begin{equation}
\log_2(\nu_n)\le \log_2(\nu_n^\mathrm{(t)})+\frac{1}{\ln(2)\nu_n^\mathrm{(t)}}(\nu_n-\nu_n^\mathrm{(t)}),
\end{equation}
where $\nu_n^\mathrm{(t)}$ is the value of $\nu_n$ at the $t$-th iteration. This linearization provides an affine upper bound on $\log_2(\nu_n)$ and, equivalently, a tight lower bound on $-\log_2(\nu_n)$:
\begin{equation}
-\log_2(\nu_n)\ge -\log_2(\nu_n^\mathrm{(t)})-\frac{1}{\ln(2)\nu_n^\mathrm{(t)}}(\nu_n-\nu_n^\mathrm{(t)}),
\end{equation}
which holds with equality when $\nu_n = \nu_n^{(t)}$. Substituting this bound into the objective function yields
\begin{align}\nonumber
&\sum_{n=1}^N\left[\log_2(\mu_n)-\log_2(\nu_n^\mathrm{(t)})-\frac{1}{\ln(2)\nu_n^\mathrm{(t)}}(\nu_n-\nu_n^\mathrm{(t)})\right]\\
=&\sum_{n=1}^N\left[\frac{\ln(\mu_n)}{\ln(2)}-\frac{\nu_n}{\ln(2)\nu_n^\mathrm{(t)}}+\frac{1}{\ln(2)}-\log_2(\nu_n^\mathrm{(t)})\right].
\end{align}
By removing constant terms, problem~\eqr{paproblem1} can be equivalently expressed as follows:
\begin{subequations}
\begin{empheq}{align}
\max_{\mathbf{p},\boldsymbol{\mu},\boldsymbol{\nu}}\quad & \sum_{n=1}^N\left[\ln(\mu_n)-\frac{\nu_n}{\nu_n^\mathrm{(t)}}\right]\\\nonumber
\textrm{s.t.} \quad & \text{(\ref{problem}f)}, \text{(\ref{problem}g)}, \text{(\ref{paproblem1}b)}, \text{(\ref{paproblem1}c)}, \text{and}\!\!\quad\!\!\! \text{(\ref{paproblem1}d)}.
\end{empheq}
\label{paproblem2}
\end{subequations}\vspace{-2mm}\\
The reformulated problem~\eqr{paproblem2} is convex and can be efficiently solved using standard convex optimization tools such as CVX~\cite{cvx}. An iterative SCA based algorithm conceived for power allocation is summarized in Algorithm~\ref{paalg}, where the initial values $\nu_n^{(0)}$ are obtained using fixed power allocation coefficients.

\begin{algorithm}[t]
\caption{SCA based algorithm for solving problem \eqr{paproblem2}}
\label{paalg}
\begin{algorithmic}[1]
\STATE Initialize $\boldsymbol{\mu}^{(0)}$ and $\boldsymbol{\nu}^{(0)}$ with the fixed power allocation.
\STATE Calculate $R^\mathrm{(0)}=\sum_{n=1}^N[\log_2(\mu_n^\mathrm{(0)})-\log_2(\nu_n^\mathrm{(0)})]$.
\STATE Set $t_{\max}$, $\epsilon$, $t=0$, and $\delta=+\infty$.
\WHILE{$\delta>\epsilon$ \AND $t<t_{\max}$}
\STATE Solve \eqr{paproblem2} to obtain $\mathbf{p}^\mathrm{(t+1)}$, $\boldsymbol{\mu}^{(t+1)}$, and $\boldsymbol{\nu}^{(t+1)}$.
\STATE $R^\mathrm{(t+1)}=\sum_{n=1}^N[\log_2(\mu_n^\mathrm{(t+1)})-\log_2(\nu_n^\mathrm{(t+1)})]$.
\STATE $\delta=\left|R^\mathrm{(t+1)}-R^\mathrm{(t)}\right|$.
\STATE $t=t+1$.
\ENDWHILE
\STATE \textbf{Output:} $\mathbf{p}^*=\mathbf{p}^\mathrm{(t)}$.
\end{algorithmic}
\end{algorithm}

The convergence of Algorithm~\ref{paalg} follows from the standard SCA framework. Specifically, the first-order approximation of the non-convex term is tight at the current point, ensuring a non-decreasing objective value over the SCA iterations. Since the achievable sum rate is upper bounded by the finite transmit power constraint, Algorithm~\ref{paalg} converges to a stationary solution of the power allocation subproblem. Together with the finite convergence of the coalitional game based user assignment and slot activation stage, the proposed two-stage framework converges to a Nash-stable user-slot configuration with stationary power allocation.
\section{Simulation Results}
In this section, simulation results are presented to demonstrate the characteristics of the proposed LCX based generalized pinching-antenna system and the effectiveness of the algorithms developed. In the simulations, users are randomly distributed within the rectangular region, while scatterers are positioned along the surrounding walls. For a given number of slots, they are evenly spaced along each cable, with an interval no smaller than half the wavelength. For comparison, the conventional fixed-antenna benchmark adopts a uniform linear array with $K \times M$ half-wavelength spaced elements at the BS located at the center of the region, matching the total number of LCX slots. Dielectric waveguide based pinching antennas are not considered as a numerical benchmark, since such architectures are primarily designed for high-frequency operation and rely on different physical structures and propagation models. The main simulation parameters are summarized in Table~\ref{parameter}. 

\begin{table}[t]
\centering
\caption{Simulation Parameters}\vspace{-1mm}
\label{parameter}
\begin{tabular}{lc}
\hline
Parameter & Value \\ \hline
Region size ($D_x$) & $50$~m \\
Attenuation constant ($\kappa$) & $0.1$~dB/m \cite{yin2024lcx} \\
Relative permittivity ($\varepsilon_r$) & $1.26$ \cite{yin2024lcx} \\
Number of scatterers ($L$) & $10$ \\
Carrier frequency ($f_c$) & $3.5$~GHz\\
Scattering path gain ($\delta_\ell$) & $\delta_\ell \sim \mathcal{CN}(0,1)$ \\
Noise power ($\sigma^2$) & $-64$~dBm\\
Convergence tolerance ($\epsilon$) & $10^{-4}$\\ \hline
\end{tabular}
\end{table}

\begin{figure}[!t]
\centering{
\subfigure[User 1 as the intended user]{\centering{\includegraphics[width=84mm]{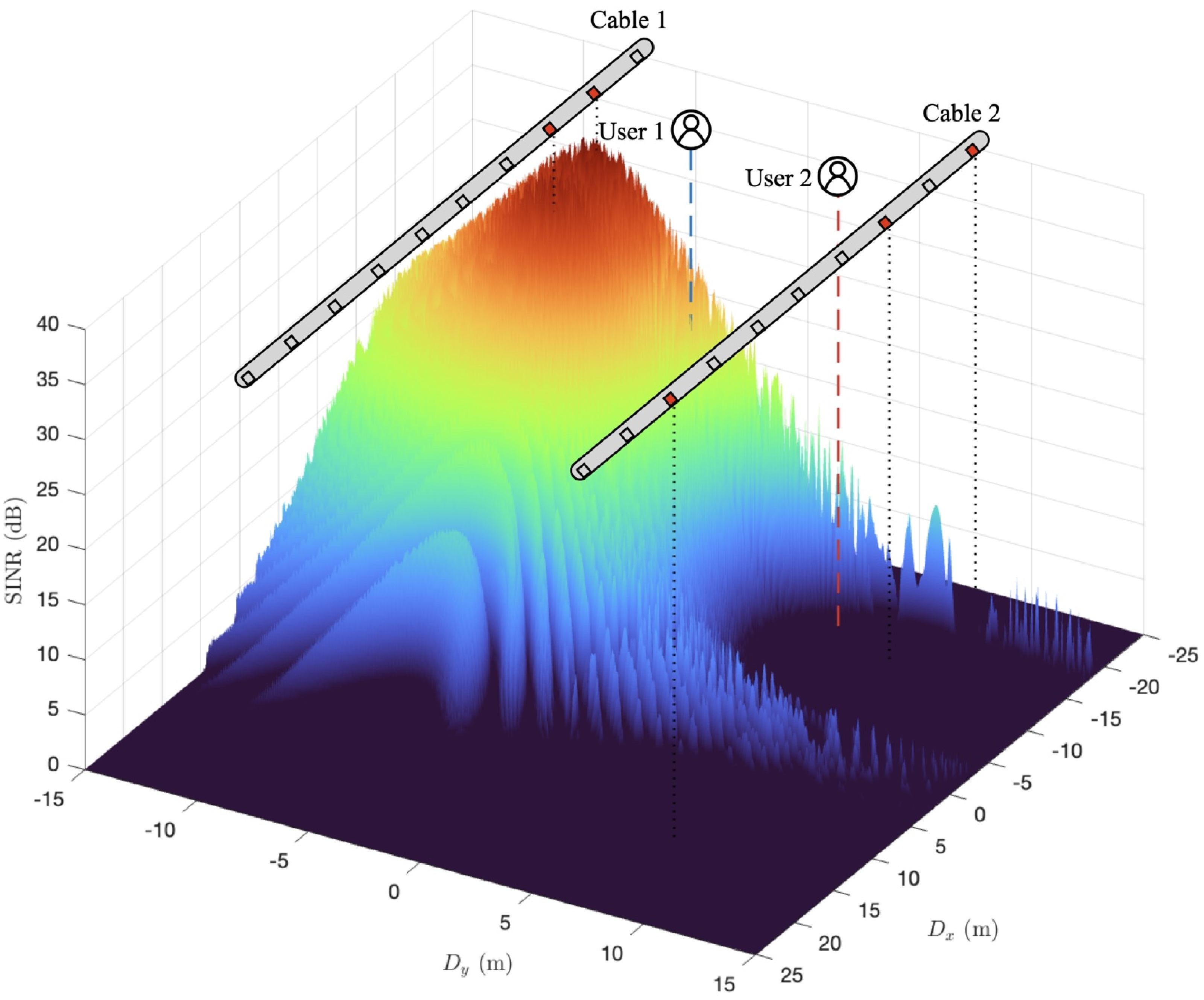}}}
\subfigure[User 2 as the intended user]{\centering{\includegraphics[width=84mm]{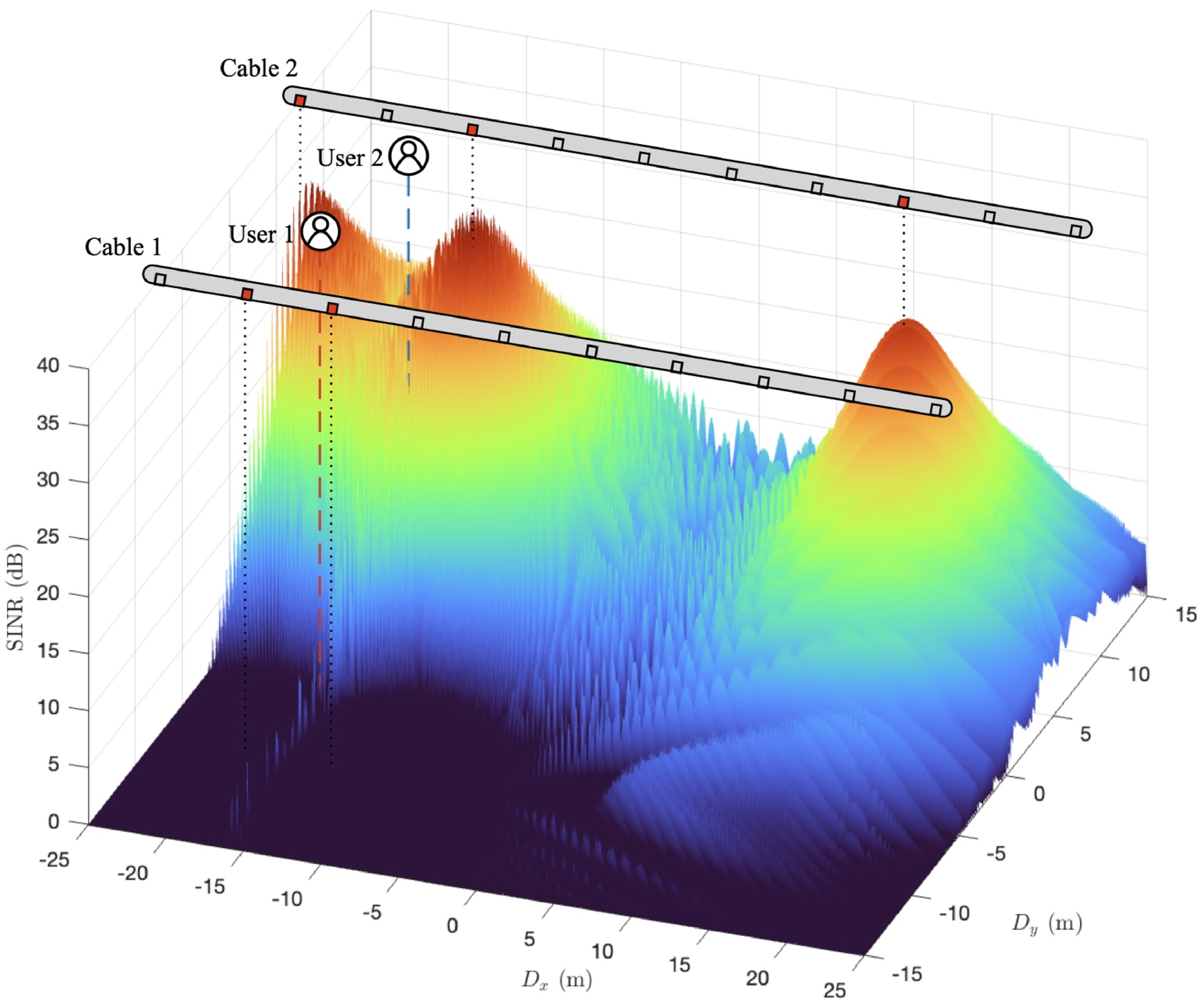}}}}
\caption{SINR distribution in the LCX based generalized pinching-antenna system, where $d=3$~m, $D_y=30$~m, $K=2$, $N=2$, $M=10$, $P_t=20$~dBm, and $R_\mathrm{min}=0.1$~bits/s/Hz.}\vspace{-4mm}
\label{result0}
\end{figure}

\fref{result0} illustrates the signal-to-interference-plus-noise ratio (SINR) distribution generated by two cables in a scatterer-free environment, where each cable serves one user. In \fref{result0}(a) and \fref{result0}(b), User 1 and User 2 are treated as the intended user, respectively, with the blue dashed line connecting each intended user to its corresponding SINR. As shown in both subfigures, the proposed slot activation strategy activates the slot closest to the intended user and additionally selects several slots on the same cable whose phases are aligned to coherently enhance the desired signal. At the same time, these selected slots are configured to destructively combine toward the unintended user (with the red dashed line indicating the corresponding SINR), thereby effectively mitigating inter-cable interference. This coordinated activation results in a strong SINR peak around the intended user while reducing the interference level experienced by the other user. 

\begin{figure}[!t]
\centering{\includegraphics[width=84mm]{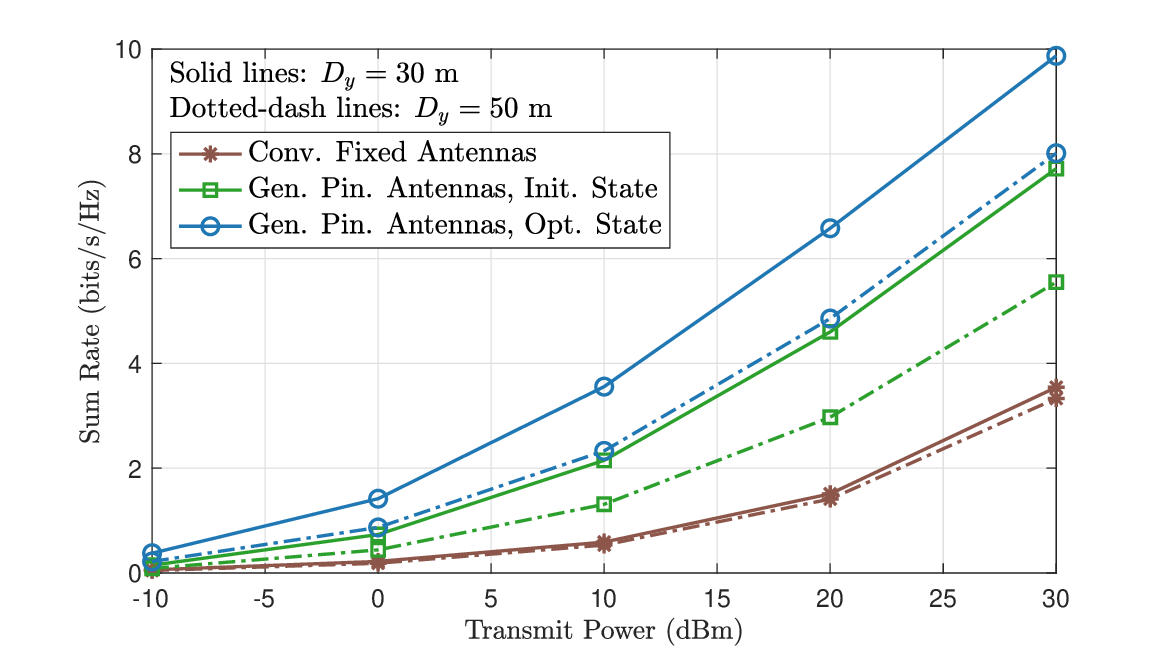}}
\caption{Impact of the transmit power on the sum rate, where $d=3$~m, $N=1$, $K=1$, $M=50$, and $R_\mathrm{min}=0.1$~bits/s/Hz.}\vspace{-4mm}
\label{result1}
\end{figure}

\fref{result1} illustrates the impact of the transmit power on the achievable sum rate for the proposed LCX based generalized pinching-antenna system. Across the entire power range, the LCX based pinching-antenna system consistently outperforms the conventional fixed-antenna architecture, illustrating the benefit of its flexible and reconfigurable radiation structure compared with static antenna deployments. A comparison between the ``Opt. State'' and ``Init. State'' curves further demonstrates the effectiveness of the proposed user assignment and slot activation algorithm. Moreover, \fref{result1} also examines the influence of the service area geometry. The LCX based system achieves significantly higher performance in elongated coverage regions, whereas its gain erodes in more square-like areas. This trend aligns with Proposition \ref{compare1}, indicating that the LCX slot radiation mechanism is particularly effective in scenarios where users are distributed along narrow corridors, tunnels, or extended deployment zones.

\begin{figure}[!t]
\centering{
\subfigure[Sum Rate]{\centering{\includegraphics[width=84mm]{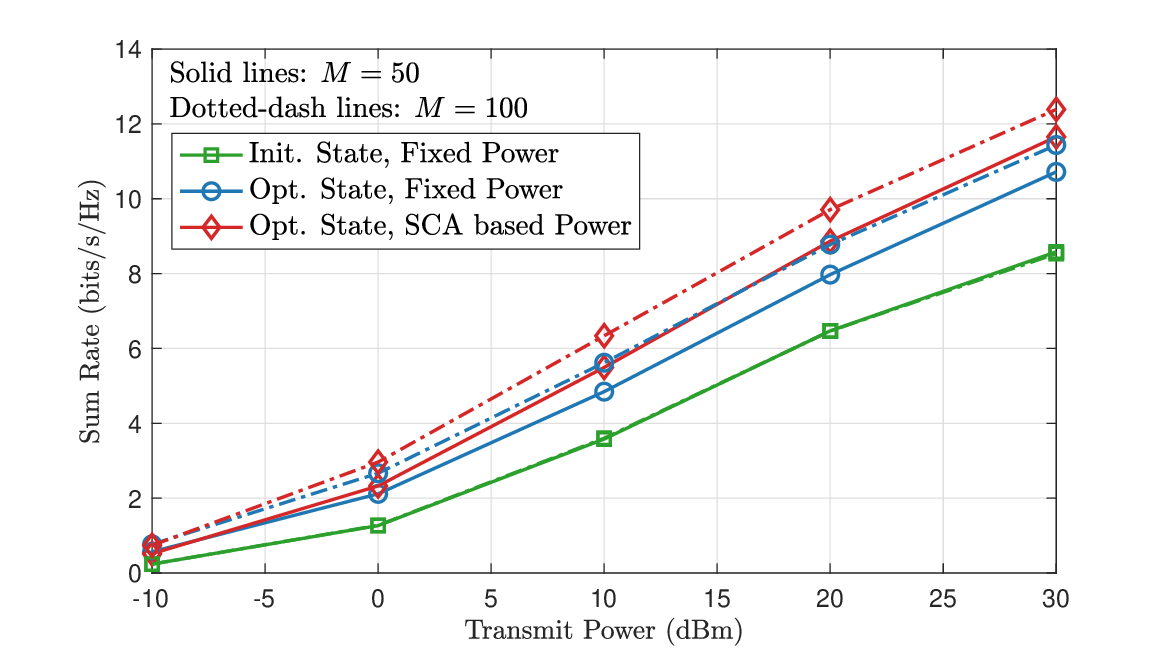}}}\vspace{-2mm}
\subfigure[Outage Probability]{\centering{\includegraphics[width=84mm]{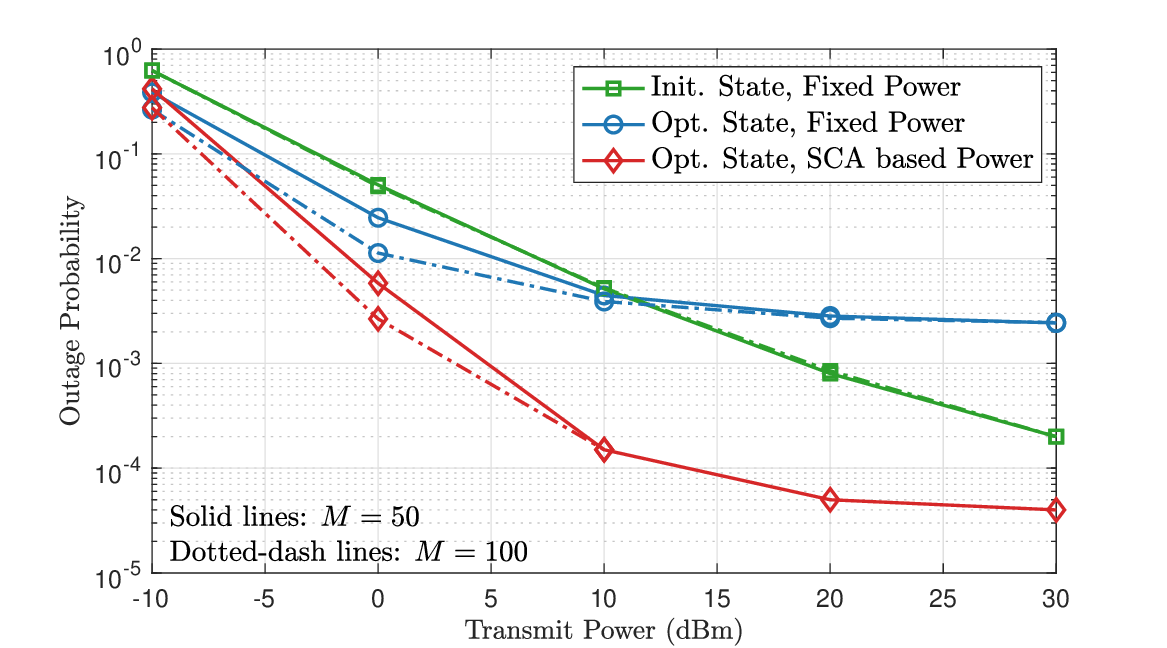}}}}\vspace{-2mm}
\caption{Impact of the transmit power on the sum rate and outage probability, where $d=3$~m, $D_y=30$~m, $K=2$, $N=2$, and $R_\mathrm{min}=0.1$~bits/s/Hz.}\vspace{-4mm}
\label{result2}
\end{figure}

\begin{figure}[!t]
\centering{
\subfigure[Sum Rate]{\centering{\includegraphics[width=84mm]{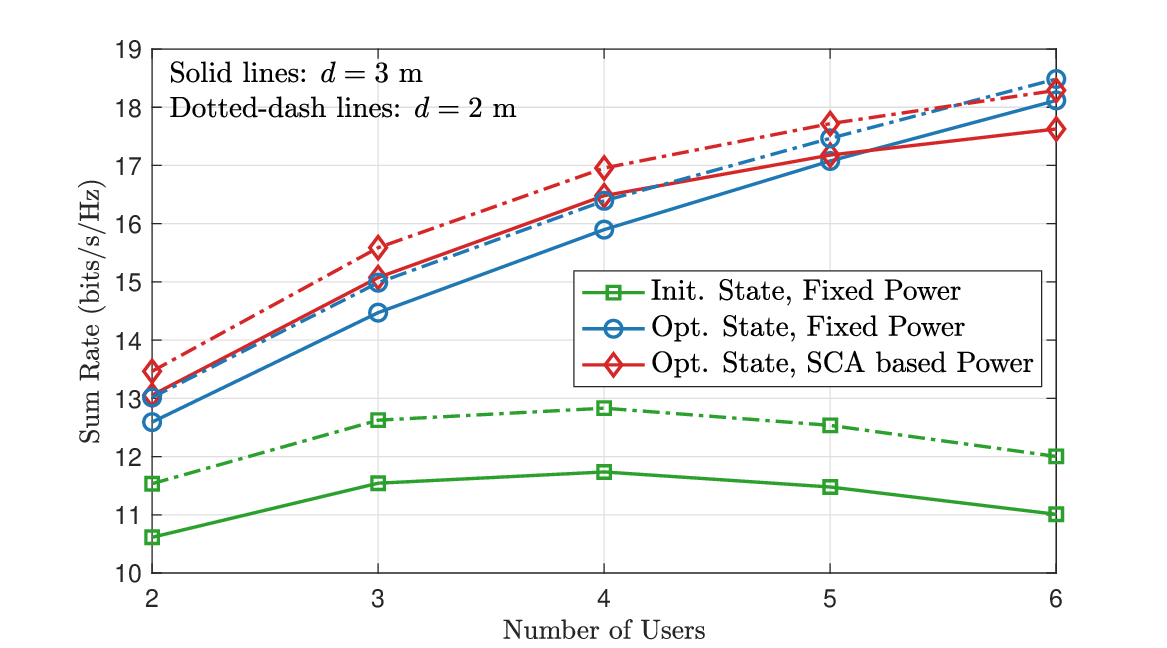}}}\vspace{-2mm}
\subfigure[Outage Probability]{\centering{\includegraphics[width=84mm]{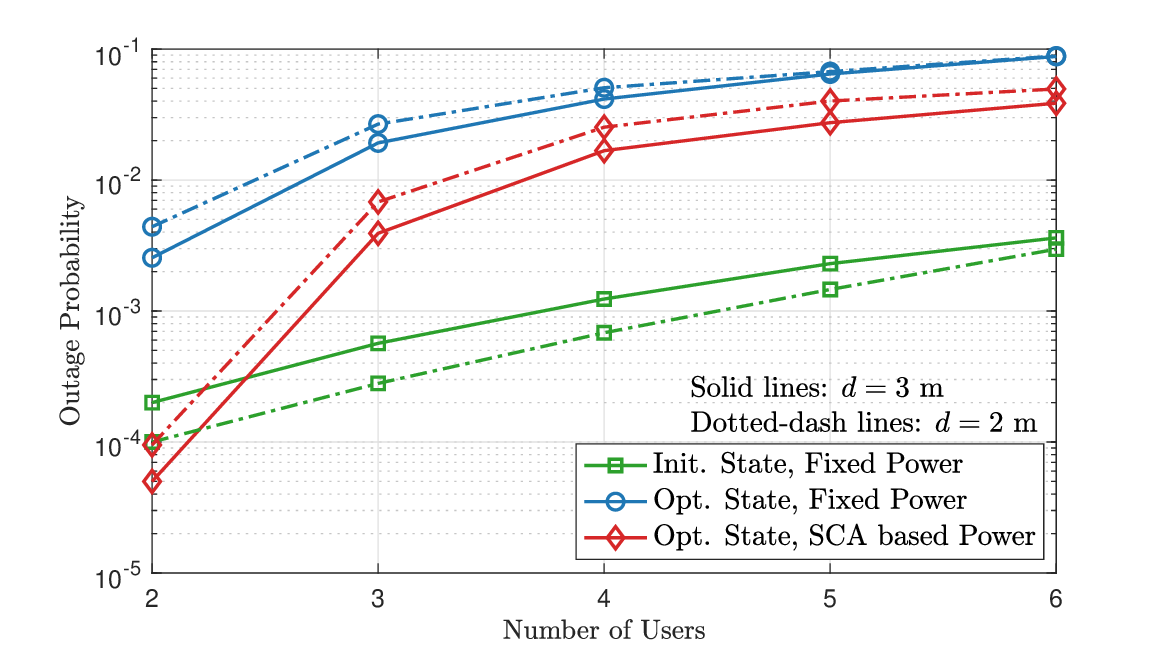}}}}\vspace{-2mm}
\caption{Impact of the number of users on the sum rate and outage probability, where $D_y=30$~m, $K=4$, $M=50$, $P_t=20$~dBm, and $R_\mathrm{min}=0.1$~bits/s/Hz.}\vspace{-4mm}
\label{result4}
\end{figure}

\fref{result2} evaluates the performance of the proposed two-stage optimization framework. As shown \fref{result2}(a), the coalitional game based user-slot association already brings about a substantial improvement in the achievable sum rate compared to the initial configuration, since cooperative grouping enables users to be reassigned to more favorable serving slots. At higher transmit power levels, this performance gain also translates into a noticeable reduction in outage probability, as illustrated in \fref{result2}(b). Building on this improved association, the subsequent SCA refinement further enhances performance by optimizing the transmit power distribution and improving the power-efficiency of the system. In addition, \fref{result4} demonstrates the impact of slot density on system performance. A higher slot density, corresponding to larger $M$, enables finer control of the effective antenna position and thus provides improved spatial flexibility, which leads to a higher achievable sum rate and lower outage probability. This observation is consistent with Remark 8, which states that increasing the slot density enhances the geometric adaptability of the system.

\begin{figure}[!t]
\centering{
\subfigure[Sum Rate]{\centering{\includegraphics[width=84mm]{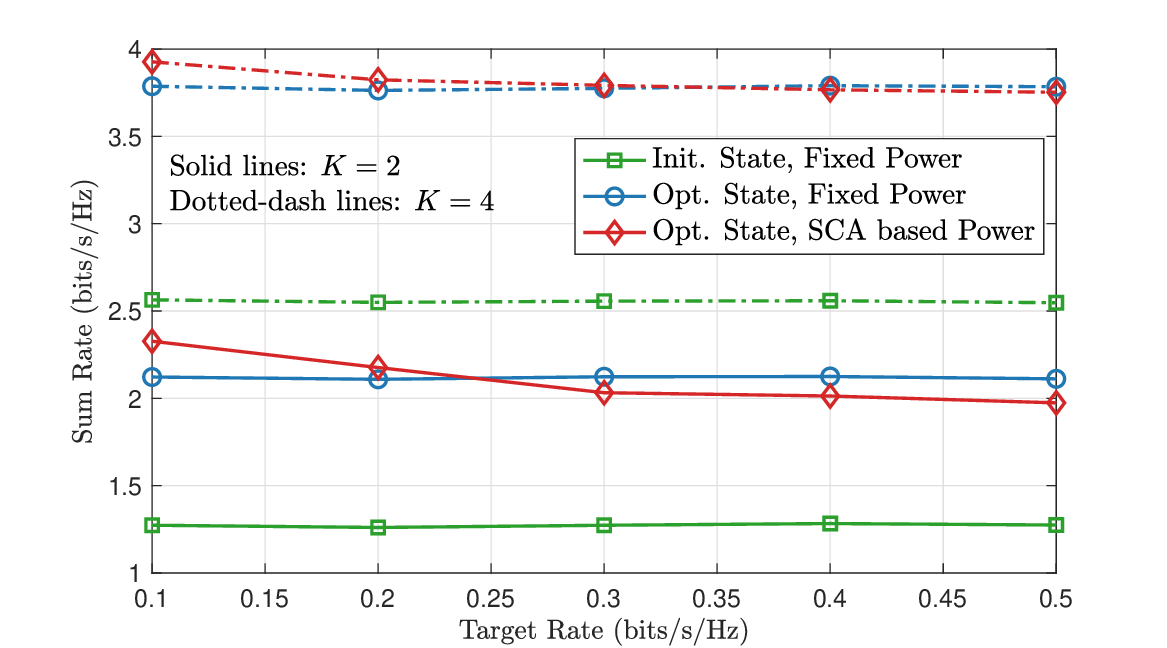}}}\vspace{-2mm}
\subfigure[Outage Probability]{\centering{\includegraphics[width=84mm]{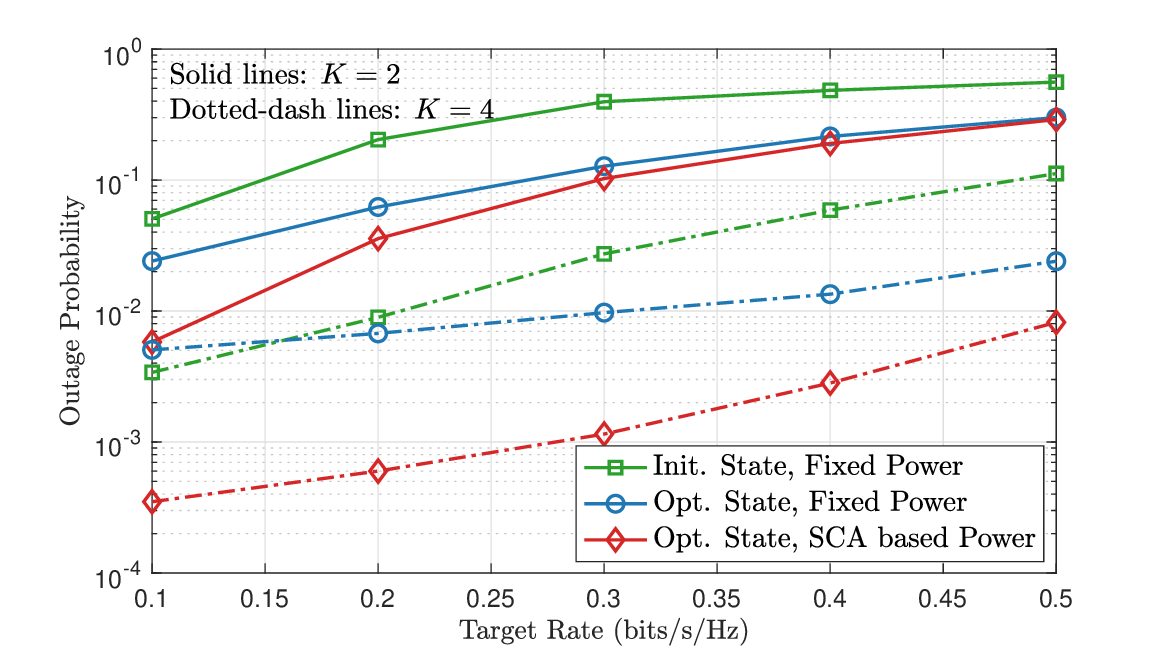}}}}\vspace{-2mm}
\caption{Impact of the target rate on the sum rate and outage probability, where $d=3$~m, $D_y=30$~m, $N=2$, $M=50$, and $P_t=0$~dBm.}\vspace{-4mm}
\label{result3}
\end{figure}

Figs. \ref{result3} and \ref{result4} collectively evaluate the performance of the proposed two-stage optimization framework. As observed from these figures, the coalitional game based algorithm improves the achievable sum rate, while also increasing the outage probability due to a more uneven rate distribution among users. Building on this association, the subsequent SCA refinement mitigates the outage by reallocating transmit power to strengthen the weakest links. When the system operates under more stringent conditions, such as high target rate requirements or a densely populated user set, the SCA based solution must allocate power more conservatively to satisfy the QoS constraints. In such scenarios, improving reliability dedicates a larger fraction of the available power budget to low-rate users and limits the achievable sum rate gain. Indeed, it may even result in a reduction of the overall throughput.

\fref{result3} further confirms the beneficial impact of the number of deployed LCXs. The use of four cables leads to a noticeable improvement in both the sum rate and outage probability compared to the two-cable case, as users gain access to more favorable elevation angles and shorter propagation distances. This observation is consistent with Remark 8, which highlights the advantage of providing additional slot diversity through multiple LCXs. Meanwhile, \fref{result4} investigates the effect of cable height. Lower cable heights result in higher sum rate because the radiated field becomes more concentrated around each cable, strengthening the intended links. This observation is consistent with Remark 4, which states that decreasing cable height enhances the elevation angle for nearby users and mitigates the interference caused by distant slots. However, the reduced height also limits the spatial reach of phase alignment, weakening the ability to mitigate inter-cable interference, which in turn leads to higher outage probability as shown in \fref{result4}(b).

\begin{figure}[!t]
\centering{\includegraphics[width=84mm]{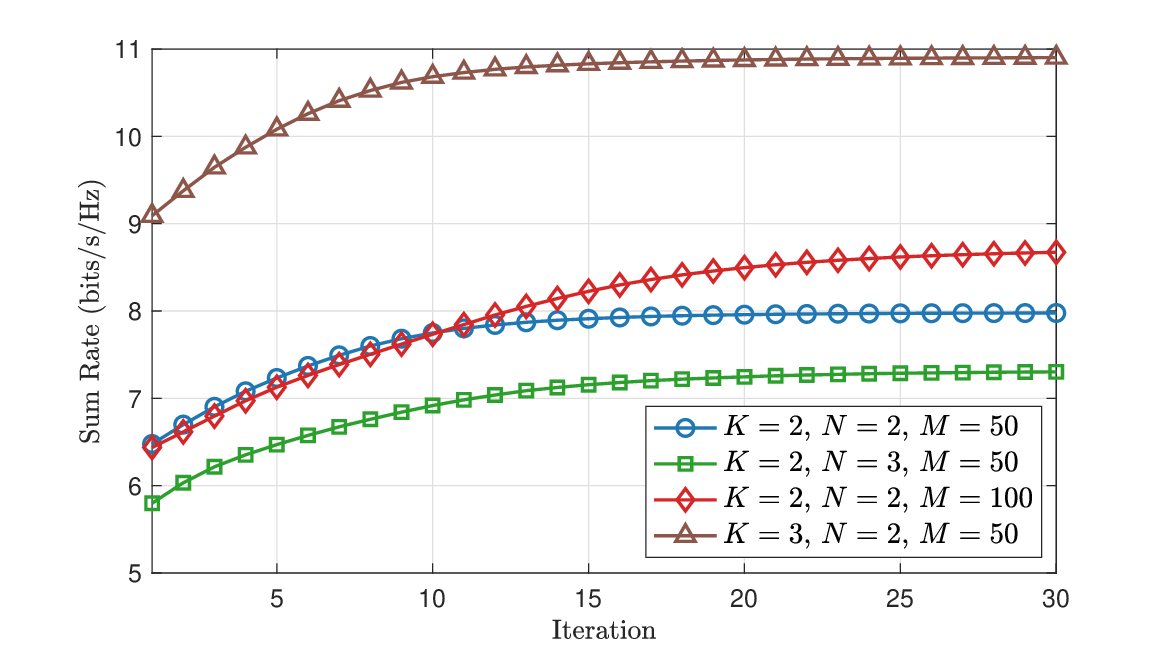}}
\caption{Convergence performance of the proposed coalitional game based algorithm, where $d=3$~m, $D_y=30$~m, $P_t=20$~dBm, and $R_\mathrm{min}=0.1$~bits/s/Hz.}\vspace{-4mm}
\label{result5}
\end{figure}

\fref{result5} illustrates the convergence behavior of the coalitional game based user assignment and slot activation algorithm. As shown in the figure, the achievable sum rate increases monotonically with each iteration and eventually converges to a stable point. This behavior is consistent with the convergence and stability properties established in the theoretical analysis. In addition, the figure reflects the fact that the algorithmic complexity is influenced by the number of cables, the number of users, and the slot density, which is in accordance with the results presented in the complexity analysis.

\begin{figure}[!t]
\centering{\includegraphics[width=84mm]{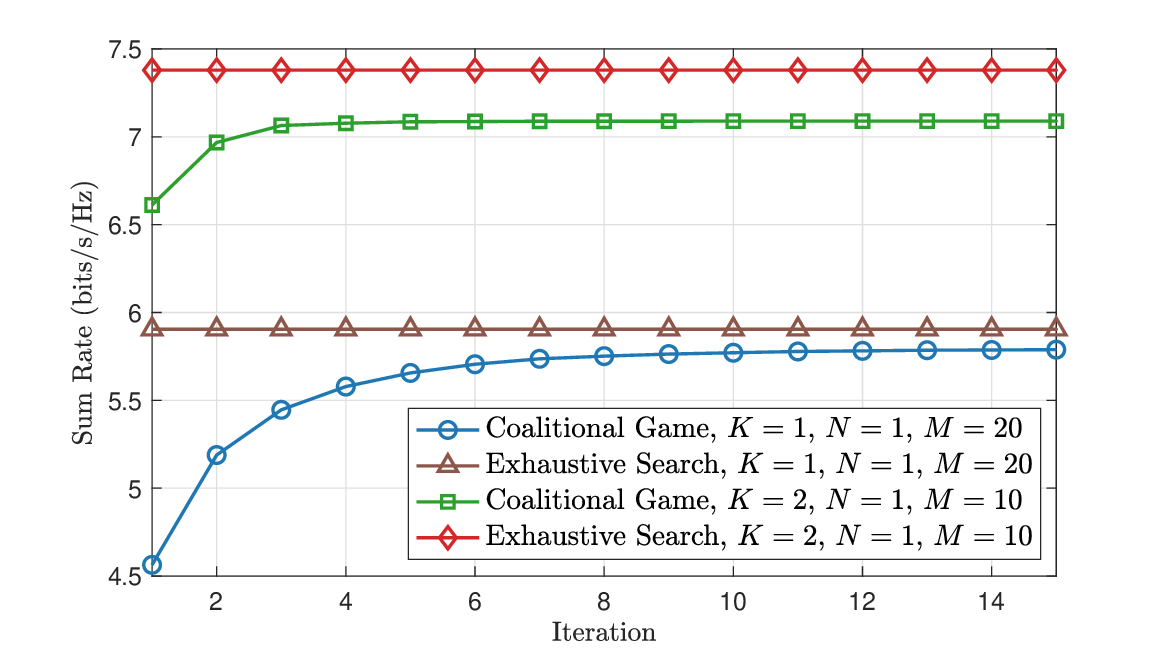}}
\caption{Comparison between the proposed coalitional game based algorithm and exhaustive search in small-scale settings, where $d=3$~m, $D_y=30$~m, $P_t=20$~dBm, and $R_{\min}=0.1$~bits/s/Hz.}\vspace{-4mm}
\label{result6}
\end{figure}

\fref{result6} compares the proposed coalitional game based algorithm using exhaustive search in small-scale settings. It can be observed that the proposed algorithm converges rapidly and achieves a sum rate close to the globally optimal solution obtained by exhaustive search. Specifically, the final sum rates of the proposed algorithm reach more than $95\%$ of the exhaustive search benchmarks in both cases. This confirms that the proposed algorithm, although suboptimal in general, can provide near-optimal performance at much lower computational complexity.
\section{Conclusions}
This paper proposed an LCX based generalized pinching-antenna architecture that enables low-frequency reconfigurable wireless communication through controllable slot activation. A two-phase channel model was established to capture both guided and radiated propagation characteristics, revealing strong local gain and rapid distance-dependent attenuation that enhance interference mitigation. A joint optimization framework was further developed to maximize the system sum rate under QoS constraints by coordinating user assignment, slot activation, and power allocation. Simulation results validated the analytical findings and demonstrated that the proposed design achieves substantial throughput and reliability improvements compared to conventional benchmarks, confirming its potential for practical low-frequency reconfigurable communication systems. As potential directions for future research, near-instantaneously adaptive forward error correction coding \cite{hanzo2002turbo}, min-max optimization techniques for enhancing rate-fairness \cite{yu2022fairness, zhu2024fairness2, zhu2025fairness1}, and Pareto optimization for characterizing multi-objective trade-offs \cite{singh2025pareto, van2026single} may be incorporated to further improve the overall system performance. Moreover, extending the proposed framework to highly dynamic scenarios through multi-timescale optimization and low-overhead reconfiguration constitutes another important research direction.
\section*{Appendix~A: Proof of Proposition~\ref{compare1}}
This comparison can be established by evaluating the lower bound of the data rate in the LCX based pinching-antenna system and the average data rate in the conventional fixed-antenna system. In the worst case, the distance between the activated slot and user $n$ is maximized, i.e., $x_\ell=\pm\tfrac{\Delta_x}{2}$ and $y_\ell=\pm\tfrac{\Delta_y}{2}$. Accordingly, the data rate of user $n$ in the LCX based system satisfies
\begin{equation}
R_n^\mathrm{LCX}\ge\underline{R}_n^\mathrm{LCX}=\log_2\!\left[1+\frac{P_n\eta^2d^2}{\sigma^2\left(\frac{\Delta_x^2}{4}+\frac{\Delta_y^2}{4}+d^2\right)^2}\right].
\end{equation}
At high transmit SNR, the above expression can be approximated as
\begin{equation}\label{rlcx}
\underline{R}_n^\mathrm{LCX}\approx\log_2\!\left(\frac{P_n\eta^2}{\sigma^2}\right)+\log_2\!\left[\frac{d^2}{\left(\frac{\Delta_x^2}{4}+\frac{\Delta_y^2}{4}+d^2\right)^2}\right].
\end{equation}

In the conventional fixed-antenna system, the expected data rate of user $n$ is given by
\begin{align}
\bar{R}_n^\mathrm{fix}&=\mathbb{E}_{\boldsymbol{\psi}_n}\left[\log_2\left(1+\frac{P_n\eta^2}{\sigma^2\left\|\boldsymbol{\psi}_n-\boldsymbol{\psi}_0\right\|^2}\right)\right]\\\nonumber
&=\frac{1}{D_xD_y}\!\int_{-\frac{D_x}{2}}^{\frac{D_x}{2}}\!\int_{-\frac{D_y}{2}}^{\frac{D_y}{2}}\!\log_2\!\!\left[1\!+\!\frac{P_n\eta^2}{\sigma^2(x^2\!+\!y^2\!+\!d^2)}\right]\!dydx.
\end{align}
Since a closed-form solution for $\bar{R}_n^\mathrm{fix}$ is difficult to obtain, an upper bound is derived by assuming that the user is uniformly distributed within a disc of radius $\tfrac{D}{2}$, where $D=\min\{D_x,D_y\}$~\cite{ding2014relay, ding2024pin}. The above integral can then be transformed as follows:
\begin{align}\nonumber
\bar{R}_n^\mathrm{fix}&\le \frac{4}{\pi D^2}\int_0^{2\pi}\int_0^{\frac{D}{2}}\log_2\left(1+\frac{P_n\eta^2}{\sigma^2(r^2+d^2)}\right)rdrd\theta\\
&=\frac{8}{D^2}\int_0^{\frac{D}{2}}\log_2\left(1+\frac{P_n\eta^2}{\sigma^2(r^2+d^2)}\right)rdr,
\label{rfix}
\end{align}
where $r=\sqrt{x_n^2+y_n^2}\le\tfrac{D}{2}$ is the radial distance from the antenna, and $\theta\in[0,2\pi)$ is the azimuthal angle in polar coordinates. According to~\cite{ding2024pin}, at high SNR,~\eqr{rfix} can be approximated as
\begin{align}\nonumber
\bar{R}_n^\mathrm{fix}\le &\log_2\left(\frac{D^2}{4}+d^2+\frac{P_n\eta^2}{\sigma^2}\right)+\log_2(e)\\\nonumber
&-\log_2\left(\frac{D^2}{4}+d^2\right)-\frac{4d^2}{D^2}\log_2\left(1+\frac{D^2}{4d^2}\right)\\\nonumber
\approx &\log_2\left(\frac{P_n\eta^2}{\sigma^2}\right)+\log_2(e)\\
&-\log_2\left(\frac{D^2}{4}+d^2\right)-\frac{4d^2}{D^2}\log_2\left(1+\frac{D^2}{4d^2}\right).
\label{rfix2}
\end{align}

By combining~\eqr{rlcx} and~\eqr{rfix2}, the rate difference can be written as follows:
\begin{align}\nonumber
&\underline{R}_n^\mathrm{LCX}-\bar{R}_n^\mathrm{fix}\\\nonumber
\ge &\log_2(d^2)-2\log_2\left(\frac{\Delta_x^2}{4}+\frac{\Delta_y^2}{4}+d^2\right)-\log_2(e)\\\nonumber
&+\log_2\left(\frac{D^2}{4}+d^2\right)+\frac{4d^2}{D^2}\log_2\left(1+\frac{D^2}{4d^2}\right)\\\nonumber
=&\log_2(d^2)-2\log_2[d^2(1+b^2)]-\log_2(e)\\\nonumber
&+\log_2[d^2(1+a^2)]+\frac{1}{a^2}\log_2(1+a^2)\\
=&\log_2(1\!+\!a^2)\!-\!2\log_2(1\!+\!b^2)\!-\!\log_2(e)\!+\!\frac{1}{a^2}\log_2(1\!+\!a^2),
\end{align}
where $a\triangleq\tfrac{D}{2d}$, and $b^2\triangleq\tfrac{\Delta_x^2}{4d^2}+\tfrac{\Delta_y^2}{4d^2}$. To satisfy $\underline{R}_n^\mathrm{LCX}\ge\bar{R}_n^\mathrm{fix}$, the following condition must hold:
\begin{align}\nonumber
&\log_2(1\!+\!a^2)\!+\!\frac{1}{a^2}\log_2(1\!+\!a^2)\ge 2\log_2(1\!+\!b^2)\!+\!\log_2(e)\\
\Rightarrow &(1+a^2)^{1+\frac{1}{a^2}}\ge e(1+b^2)^2.
\end{align}
This inequality can be equivalently expressed as
\begin{equation}
\left(1+\frac{D^2}{4d^2}\right)^{\left(1+\frac{4d^2}{D^2}\right)}\ge e\!\left(1+\frac{\Delta_x^2+\Delta_y^2}{4d^2}\right)^2.
\end{equation}
Defining
\begin{equation}
f(a)=\log_2(1\!+\!a^2)\!+\!\frac{1}{a^2}\log_2(1\!+\!a^2),
\end{equation}
its derivative is given by
\begin{align}\nonumber
f'(a)&=\frac{2}{a\ln(2)}-\frac{2\ln(1+a^2)}{a^3\ln(2)}\\
&=\frac{2}{\ln(2)a^3}[a^2-\ln(1+a^2)].
\end{align}
Since $a^2 \ge 0$ and $\ln(1+a^2) \le a^2$, it follows that $f'(a) \ge 0$. Therefore, $\underline{R}_n^{\mathrm{LCX}} - \bar{R}_n^{\mathrm{fix}}$ is monotonically increasing with $\tfrac{D}{d}$, and this proof is completed.\QEDA
\section*{Appendix~B: Proof of Proposition~\ref{compare2}}
Consider the case in which each cable or waveguide has only one activated antenna, and NLoS components are neglected. In this scenario, the phase terms have no impact on the received signal, and the data rate of user $n$ in the LCX based pinching-antenna system can be expressed as follows:
\begin{align}\nonumber
R_n^\mathrm{LCX}&=\log_2\left(1+\frac{P_n|h_{k,m,n}|^2}{P_n|h_{k',m',n}|^2+\sigma^2}\right)\\
&=\log_2\!\left[1\!+\!\frac{\mu_{k,m,n}^2\sin^2(\phi_{k,m,n})}{\mu_{k',m',n}^2\sin^2(\phi_{k',m',n})+\frac{\sigma^2}{P_n\eta^2}}\right],
\end{align}
where
\begin{equation}
\mu_{k,m,n}=\frac{10^{-\frac{\kappa}{20}\|\boldsymbol{\psi}_{k,0}-\boldsymbol{\psi}_{k,m}^\mathrm{slot}\|}}{\|\boldsymbol{\psi}_n-\boldsymbol{\psi}_{k,m}^\mathrm{slot}\|}.
\end{equation}
It is assumed that all users transmit at equal power $P_n$. Similarly, the data rate of user $n$ under the conventional pinching-antenna channel model is given by
\begin{equation}
R_n^\mathrm{Pin}=\log_2\left(1+\frac{\mu_{k,m,n}^2}{\mu_{k',m',n}^2+\frac{\sigma^2}{P_n\eta^2}}\right).
\end{equation}
Since $\mu_{k,m,n}^2>0$ and $\mu_{k',m',n}^2>0$ always hold, the LCX based channel model yields a higher data rate than the conventional pinching-antenna channel model if and only if
\begin{align}\nonumber
&R_n^\mathrm{LCX}\ge R_n^\mathrm{Pin}\\\nonumber
\Rightarrow &\frac{\sin^2(\phi_{k,m,n})}{\mu_{k',m',n}^2\sin^2(\phi_{k',m',n})+\frac{\sigma^2}{P_n\eta^2}}\ge \frac{1}{\mu_{k',m',n}^2+\frac{\sigma^2}{P_n\eta^2}}\\
\Rightarrow & \sin^2(\phi_{k,m,n})\!-\!\sin^2(\phi_{k',m',n})\ge\gamma_{k',m',n}[1\!-\!\sin^2(\phi_{k,m,n})],
\end{align}
where $\gamma_{k',m',n}\triangleq\tfrac{\sigma^2}{P_n\eta^2\mu_{k',m',n}^2}$. The above inequality can be rearranged as follows:
\begin{align}\nonumber
&R_n^\mathrm{LCX}\ge R_n^\mathrm{Pin}\\\nonumber
\Rightarrow &\sin^2(\phi_{k,m,n})(1\!+\!\gamma_{k',m',n})\ge\gamma_{k',m',n}\!+\sin^2(\phi_{k',m',n})\\
\Rightarrow &\sin^2(\phi_{k,m,n}) \ge \frac{\gamma_{k',m',n}\!+\sin^2(\phi_{k',m',n})}{1+\gamma_{k',m',n}}.
\end{align}

At high SNR, $\frac{\sigma^2}{P_n\eta^2}\to 0$, and the corresponding data rates can be approximated as
\begin{equation}
\tilde{R}_n^\mathrm{LCX}\approx\log_2\!\left[1+\frac{\mu_{k,m,n}^2\sin^2(\phi_{k,m,n})}{\mu_{k',m',n}^2\sin^2(\phi_{k',m',n})}\right],
\end{equation}
and
\begin{equation}
\tilde{R}_n^\mathrm{Pin}\approx\log_2\!\left(1+\frac{\mu_{k,m,n}^2}{\mu_{k',m',n}^2}\right).
\end{equation}
In this context,
\begin{align}\nonumber
\tilde{R}_n^\mathrm{LCX}\ge \tilde{R}_n^\mathrm{Pin}\Rightarrow & \frac{\mu_{k,m,n}^2\sin^2(\phi_{k,m,n})}{\mu_{k',m',n}^2\sin^2(\phi_{k',m',n})}\ge \frac{\mu_{k,m,n}^2}{\mu_{k',m',n}^2}\\
\Rightarrow & \sin^2(\phi_{k,m,n})\ge \sin^2(\phi_{k',m',n}),
\end{align}
as $\mu_{k,m,n}^2>0$ and $\mu_{k',m',n}^2>0$. Since $0<\phi\le\tfrac{\pi}{2}$ and $\sin(\phi)$ is a monotonically increasing function, the above inequality can be equivalently expressed as follows:
\begin{equation}
\tilde{R}_n^\mathrm{LCX}\ge \tilde{R}_n^\mathrm{Pin}\Leftrightarrow\phi_{k,m,n}\ge \phi_{k',m',n}.
\end{equation}
This condition is further equivalent to
\begin{equation}
\tilde{R}_n^\mathrm{LCX}\ge \tilde{R}_n^\mathrm{Pin}\Leftrightarrow \left\|\boldsymbol{\psi}_n-\boldsymbol{\psi}_{k,m}^\mathrm{slot}\right\|\le \left\|\boldsymbol{\psi}_n-\boldsymbol{\psi}_{k',m'}^\mathrm{slot}\right\|,
\end{equation}
and this proposition is proven.\QEDA
\bibliographystyle{IEEEtran}
\bibliography{KaidisBib}
\end{document}